\documentclass[sigconf]{acmart}

\AtBeginDocument{%
  \providecommand\BibTeX{{%
    \normalfont B\kern-0.5em{\scshape i\kern-0.25em b}\kern-0.8em\TeX}}}

\copyrightyear{2026}
\acmYear{2026}
\setcopyright{cc}
\setcctype{by}
\acmConference[ASSETS '26]{The 28th International ACM SIGACCESS Conference on Computers and Accessibility}{October 25--28, 2026}{Vila Nova de Gaia, Portugal}
\acmBooktitle{The 28th International ACM SIGACCESS Conference on Computers and Accessibility (ASSETS '26), October 25--28, 2026, Vila Nova de Gaia, Portugal}
\acmDOI{10.1145/3797867.3828994}
\acmISBN{979-8-4007-2521-0/2026/10}

\usepackage{tabularx}
\usepackage{multirow}
\usepackage{graphicx}
\usepackage{subcaption}

\newcommand{\softwarename}{VisionPulse}
\newcommand{\numparticipants}{12}

\begin{document}

\title[\softwarename{}: A Virtual Reality System Enabling Accessible Discovery and Navigation for Blind and Low Vision Users]{\softwarename{}: A Virtual Reality System Enabling Accessible Discovery and Navigation for Blind and Low Vision Users}

\author{Samuel Martin}
\affiliation{
  \institution{Arizona State University}
  \city{Tempe}
  \state{Arizona}
  \country{USA}
}
\email{simart12@asu.edu}

\author{Pooyan Fazli}
\affiliation{
  \institution{Arizona State University}
  \city{Tempe}
  \state{Arizona}
  \country{USA}
}
\email{pooyan@asu.edu}

\author{Hasti Seifi}
\affiliation{
  \institution{Arizona State University}
  \city{Tempe}
  \state{Arizona}
  \country{USA}
}
\email{hasti.seifi@asu.edu}

\begin{abstract} 
Free exploration is an important aspect of many engaging virtual reality (VR) experiences, yet remains largely inaccessible to blind and low vision (BLV) users due to its reliance on visual feedback. Existing approaches support BLV navigation through prebuilt menus of environment and audio beacons, but offer limited support for free-form discovery. We present \softwarename{}, an accessible VR system that enables BLV users to explore virtual environments through natural head and hand movements, combined with auditory, haptic, and text-to-speech feedback. \softwarename{} introduces a discovery-driven approach that allows users to progressively uncover regions and objects, alongside navigation support through waypoint guidance and object localization via responsive audio and orientation-based haptics. A study with 12 BLV participants showed a strong preference for \softwarename{}’s discovery-based exploration and multimodal feedback, without negatively impacting task performance or perceived workload. Our findings underscore the importance of accessible, free-form VR experiences, and contribute insights for inclusive VR design.
\end{abstract}




\begin{teaserfigure}
  \includegraphics[width=\textwidth]{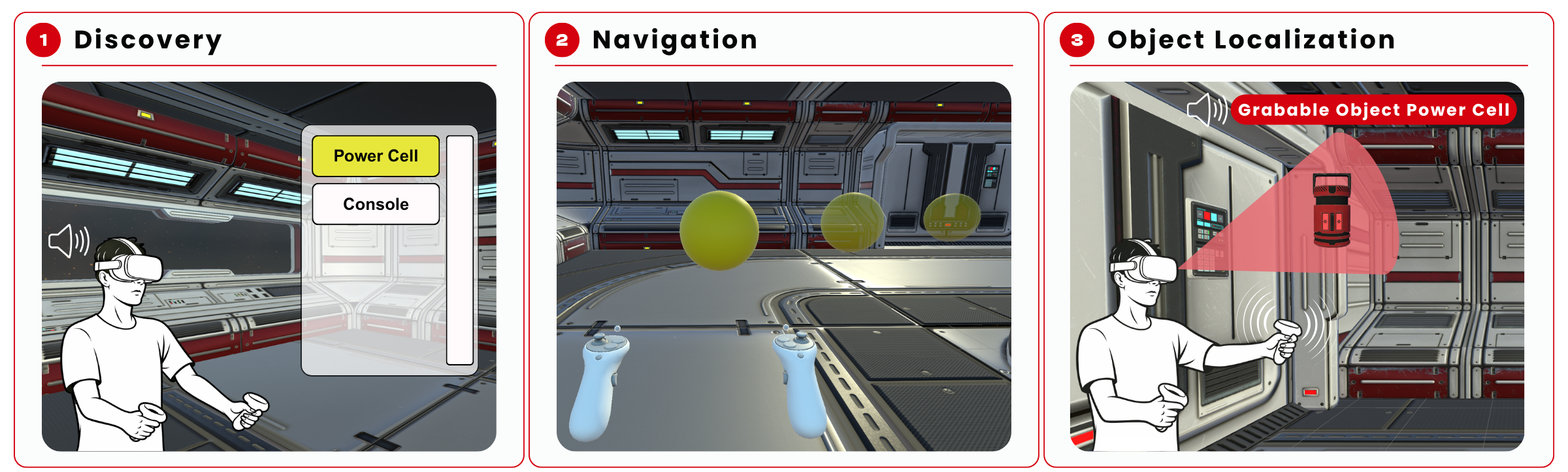}
  \caption{\softwarename{} enables blind and low vision (BLV) users to discover and navigate virtual environments. (1) Users explore the environment via head movements and open the \emph{discovery menu} to select objects of interest. 
  (2) The system then guides navigation using a sequence of waypoints (yellow spheres). (3) To support orientation, spatial audio tones varying in pitch and volume, and controller vibrations provide directional guidance and help precisely locate nearby waypoints and objects in 3D space. When a controller touches the object, text-to-speech then indicates the object is graspable.}
  \label{fig:teaser}
\end{teaserfigure}


\maketitle

\section{Introduction}
An estimated 2.2 billion people globally live with some form of vision impairment or blindness~\cite{WHO2026Blindness}, highlighting a significant accessibility gap in emerging technologies. Despite this, many digital experiences, especially immersive ones, remain designed primarily for sighted users. Virtual Reality (VR) is rapidly evolving from a niche technology into a medium transforming how people work, learn, socialize, and play. The global VR market is projected to grow from USD 26.71 billion in 2026 to USD 171.33 billion by 2034~\cite{vrMarket}. In many VR experiences, sighted users can freely explore and discover environments, which foster user agency, active participation, and engagement. However, for people with blindness or low vision (BLV), such exploration remains largely inaccessible due to reliance on visual interfaces and feedback.

Prior work has primarily focused on guided VR navigation for BLV users, with only limited efforts toward discovery-based interaction in video games. Some studies rely on prebuilt menus of virtual objects~\cite{trewin2008powerup} paired with audio beacons (i.e., spatial audio from object locations) to locate objects~\cite{lan2024speed, mendes2025exploring}. While these approaches can support efficient navigation, they constrain user curiosity and exploration. Recent work highlights the importance of supporting discovery in 3D video games, proposing a system that enables BLV players to scan the virtual space using a joystick-controlled beam and receive spatial audio cues naming encountered objects~\cite{nair2024surveyor,nair2021navstick}.
However, this approach does not fully leverage the embodied affordances of VR, such as head and body movements and multimodal audio-haptic feedback. As a result, exploration and object localization can become inefficient and physically demanding, particularly in environments where architectural elements (e.g., ramps, stairs) and objects are distributed across varying heights and positions in 3D space.

To address this gap, we present \softwarename{}, an accessible VR system that supports BLV users in discovery, navigation, and object localization through embodied interaction and responsive audio and haptic feedback (Figures~\ref{fig:teaser}). \softwarename{} enables users to explore VR environments by naturally moving their head, progressively revealing regions and objects through movement and interaction. Discovered elements are dynamically added to the \emph{discovery menu} and conveyed to users via text-to-speech (TTS) feedback. 
The system supports navigation to distant sections and objects in the \emph{discovery menu} through waypoints, while enabling interaction with nearby targets via multimodal feedback. Users receive continuous audio cues responsive to head orientation and haptic feedback aligned with hand orientation and  distance, helping them orient toward points of interest and localize them in 3D space for interaction. These cues adapt to user movement, providing intuitive, physically grounded guidance. Together, they create a unified interaction model for both open-ended exploration and efficient navigation.

To evaluate \softwarename{}, we conducted a user study with \numparticipants{} BLV participants addressing two research questions: (RQ1) How do BLV users experience \softwarename{}’s discovery and multimodal feedback mechanisms? (RQ2) How does free-form exploration affect task performance and workload compared to a pre-discovered environment? We used a $2\times3$ within-subjects design comparing two exploration types (\emph{prebuilt menu} vs. \softwarename{}’s \emph{discovery}) and three feedback modalities (\emph{audio beacon}, \emph{responsive audio beacon}, and \emph{responsive audio beacon with haptics}). Participants completed game-like tasks across six VR conditions and provided feedback through ratings and interviews. 
Qualitative results showed a strong preference for discovery-based exploration, with participants reporting greater autonomy, engagement, and sense of exploration. 
Multimodal feedback, especially responsive audio with haptics, was consistently favored for improving users' confidence and sense of orientation. While discovery led to longer paths and more exploration, it did not significantly increase perceived workload or completion time, suggesting increased exploration without reduced efficiency. This work makes two primary contributions:

\begin{itemize}
    \item\softwarename{}, an embodied VR system that supports BLV users in free-form discovery of environments via head movements, guides navigation to distant targets via waypoints and responsive audio, and enables precise localization and interaction with nearby objects through responsive audio, haptic, and TTS feedback.
    \item Results on BLV users' experience, workload, and task performance with \softwarename{}'s interaction techniques.
\end{itemize}

To support future research and the design of accessible VR experiences for BLV users, we make \softwarename{} open source at \url{https://github.com/teal-lab/VisionPulse}.

\section{Related Work}
We review prior work on accessibility in games and VR for BLV users, focusing on research that supports discovery, navigation, and object localization.

\subsection{\textbf{Accessible Discovery in Virtual Environments}}
Prior work has highlighted persistent challenges faced by BLV users when exploring virtual environments~\cite{andrade2019playing, martinez2024playing, ran2025how}. 
Some solutions leverage pre-structured menus that allow users to survey virtual objects through linear lists~\cite{trewin2008powerup, matsuo2016audible}. While effective for conveying information, such methods can limit user enjoyment from self-directed exploration and discovery. 
More recent research has shifted toward supporting free-form exploration in VR. For instance, Collins et al. pair BLV participants with sighted guides to facilitate VR exploration and interaction~\cite{collins2023guide}. However, reliance on human assistance may not always be practical and can reduce users’ independence and privacy~\cite{akter2020uncomfortable}. Other work has explored automated techniques to enhance environmental awareness. Ji et al. introduce a proximity-based approach that defines awareness zones around the user, triggering audio feedback when nearby entities are detected~\cite{ji2022vrbubble}. Yet, this approach uses predefined spatial boundaries and passively delivers information within a fixed radius, which can be overwhelming in complex VR environments with many objects and users~\cite{ji2022vrbubble}. 

Alternative approaches have investigated user-initiated cues for environment discovery. Some used echolocation in virtual environments, where users rely on self-generated sounds (e.g., clicks, footsteps) and their reflections to infer spatial properties. These studies suggest that echolocation can help BLV users develop a general understanding of environmental layout, but it remains less effective for perceiving finer spatial details such as room geometry~\cite{andrade2018echo, andrade2021echolocation}. 
Another approach, such as NavStick~\cite{nair2021navstick} and Surveyor~\cite{nair2024surveyor}, enable users to actively probe their surroundings by casting a virtual beam using a joystick. As the beam intersects with objects, users receive verbal labels, allowing them to query the environment on demand and fostering a strong sense of freedom and enjoyment among BLV users. 
\softwarename{} advances this discovery-driven paradigm by leveraging VR’s embodied affordances, enabling BLV users to naturally reveal spatial information and objects through head movements, thereby improving the efficiency of VR exploration.

\subsection{\textbf{Navigation to Targets}}
Prior work has extensively explored navigation guidance for BLV users across both physical and virtual contexts. A large body of literature focuses on real-world navigation, including approaches that leverage smartphones to provide on-the-go assistance~\cite{kubota2024snap, india2021vstroll}, as well as systems that allow BLV users to preview virtual routes before physical traversal~\cite{guerreiro2017virtual, guerreiro2020virtual, lahav2022virtual, thevin2020xroad, connors2014virtual}. In parallel, other research has examined navigation within screen-based 3D environments, investigating how BLV users orient themselves in virtual spaces~\cite{nair2024surveyor, facanha2020o&m}.

To support BLV orientation and target-directed movement, many systems rely on auditory feedback. For example, Piçarra et al. employ structured verbal instructions to guide participants through a labyrinth~\cite{pi2023evaluating}. Other approaches use audio beacons, spatialized sound cues that indicate the target's direction, to support virtual navigation~\cite{walker2003effect,walker2006navigation,lastofus2_2020,nair2024surveyor}. Others encode navigation-relevant information through dynamic audio variations~\cite{smith2018rad,miura2023exploration}. For instance, RAD, a car racing game designed for BLV users, conveys speed, trajectory, and upcoming turns through continuous auditory cues that reflect direction, sharpness, length, and timing~\cite{smith2018rad}. Building on this, \softwarename{} proposes responsive audio beacons that dynamically adapt to head movement, supporting BLV user orientation toward navigation waypoints.

Prior work has also investigated VR locomotion techniques with BLV users. Several studies have explored white cane-based techniques that enable simultaneous physical walking and virtual movement~\cite{siu2020virtual, zhao2018enabling, shrestha2025virtual}. Others have examined BLV experiences for existing VR locomotion techniques. One study reports that joystick-based locomotion provides higher perceived safety, precision, and ease of use compared to omni-directional treadmills and walk-in-place locomotion~\cite{kreimeier2020blindwalkvr}. Similarly, another study suggests that joystick locomotion is favored for its simplicity and familiarity, but also noted diverse BLV user preferences for the locomotion technique~\cite{ribeiro2024investigating}. In \softwarename{}, we adopt joystick-based locomotion and discuss its implications on BLV experiences in VR.

\subsection{\textbf{Virtual Object Awareness and Localization}}
Prior work has explored accessible methods for BLV users to localize and interact with virtual objects using audio and haptic feedback~\cite{mendes2025exploring, wedoff2019virtual}. Guerreiro et al. proposed a design space for nonvisual auditory representations to convey spatial information, object properties, and interactions~\cite{guerreiro2023the}. Furtado et al. developed an accessible virtual boxing game using layered audio cues for opponent actions and feedback~\cite{furtado2025designing}. Lança et al. further investigated object localization in VR through their Speed of Light game, comparing speech feedback, sonification, and a 2D grid-based auditory representation~\cite{lan2024speed}, while Virtual Showdown used concise verbal cues to indicate the position of a moving ball in a virtual tennis scenario~\cite{wedoff2019virtual}. Beyond gaming, SceneVR enabled BLV users to inspect and localize virtual objects by streaming a VR scene to a mobile device and providing verbal descriptions of touched objects through spatial audio~\cite{kneitmix2025from}. More recent approaches leveraged AI-driven audio description techniques~\cite{kulkarnivideosavi,videopasta} to automatically generate verbal and spatialized descriptions of virtual scenes and objects~\cite{killough2025vrsight}. 
\softwarename{} uses brief TTS-based audio descriptions to convey information about the environment and objects, highlighting discovered items and waypoints along with their interactivity status, while responsive audio beacons orient users toward targets.

Haptic feedback has also been widely explored for object localization and interaction. Several studies rely on custom hardware, including haptic canes~\cite{zhao2018enabling,siu2020virtual, schloerb2010blindaid} or handheld devices~\cite{schneider2018dualpanto, morelli2018vi}. However, such specialized hardware is often not widely accessible and may introduce additional cost and maintenance burdens. Other work has leveraged vibration feedback available in commodity devices. For example, Wedoff et al. modulate vibration amplitude based on proximity to a moving virtual ball in a table tennis game~\cite{wedoff2019virtual}. Wald et al. propose two-handed localization techniques 
which compute vibration feedback relative to the user’s hand position to target~\cite{wald2025spatial}. This technique primarily encodes distance through vibration intensity and have not been evaluated with BLV users. 
\softwarename{} uses vibration feedback to support object localization, but emphasizes the controller’s angular orientation relative to target objects while still incorporating distance cues. This approach allows users to directly align their hand with targets, enabling effective spatial search in 3D environments where objects vary in height and position.

\begin{figure*}[htbp] 
    \centering
    \includegraphics[width=0.9\textwidth]{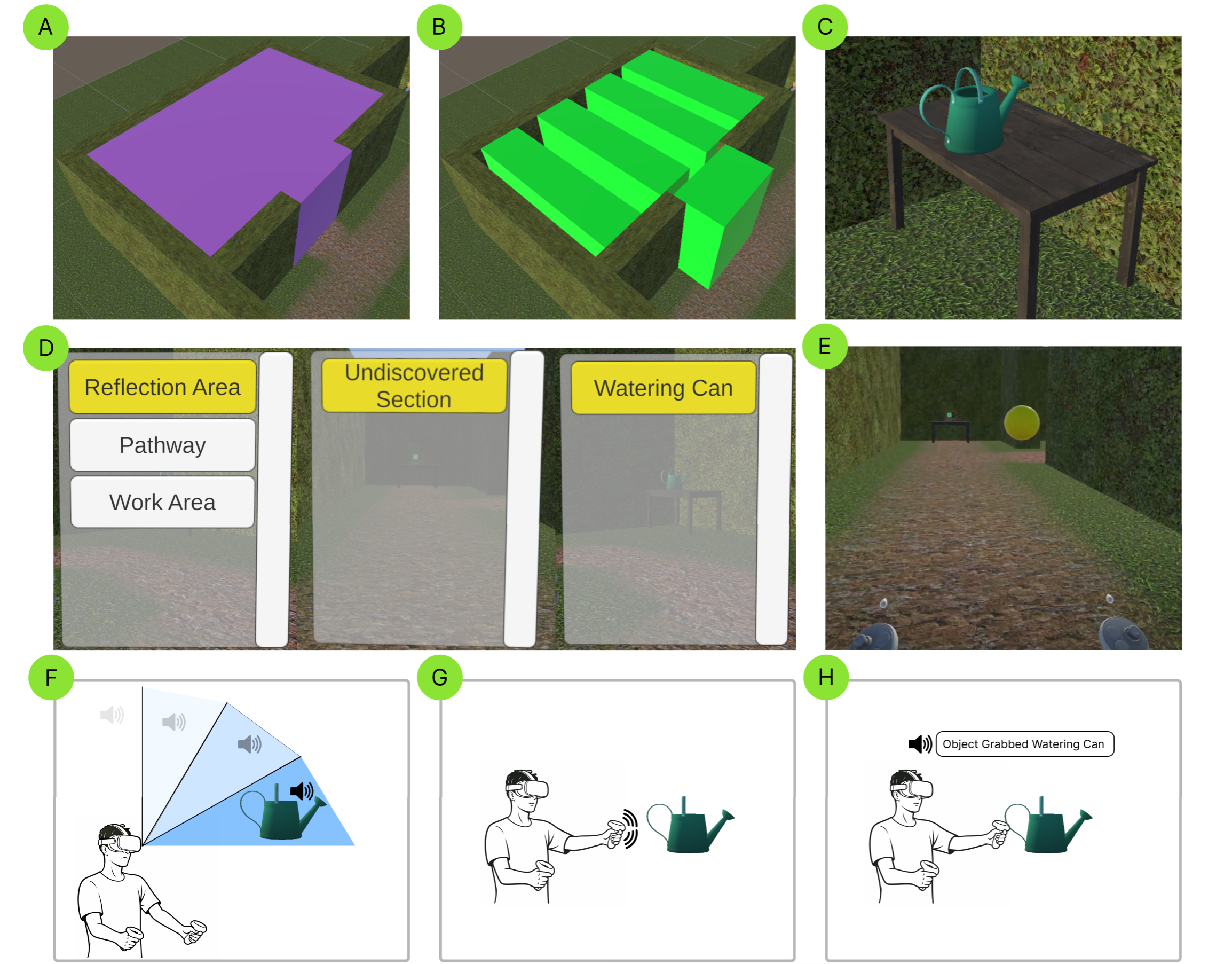}
    \caption{Users begin in the discovery phase, encountering a new portion of the virtual environment with (A) a region, (B) segmented sections (shown with gaps for illustration), and (C) objects (e.g., a watering can). Users can then open the \emph{discovery menu} (D) to highlight a region and select either an object or an undiscovered section within the region. The system then transitions to the navigation phase, guiding the user via waypoints (yellow spheres) (E). During navigation, both waypoints and the target object provide guidance using the same auditory and haptic cues: a spatialized hum that varies in volume and pitch based on head orientation (F), and controller vibrations that increase as the controller aligns more directly with the target (G). When the controller contacts the object, text-to-speech (TTS) confirms the object is graspable and confirms again upon successful grasp (H). The system returns to the discovery phase, repeating the interaction cycle.
    }
    \label{fig:workflow}
\end{figure*}

\section{\softwarename{}}
\softwarename{} is an accessible VR system, developed in the Unity Game Engine, that combines a structured virtual world setup, embodied interaction, and multimodal feedback to enable BLV users to discover virtual spaces, navigate to distant targets, and precisely locate and interact with nearby items (Figure~\ref{fig:workflow}). 
Unlike prior approaches such as Surveyor~\cite{nair2024surveyor} and NavStick~\cite{nair2021navstick}, which require users to explicitly query the environment through joystick-based beam casting, \softwarename{} leverages the embodied affordances of immersive VR to support continuous discovery, navigation, and interaction through natural head and hand movements. Head orientation directs exploration by progressively revealing regions and objects, responsive audio beacons guide users toward regions and targets, and controller haptics support precise hand-based object localization. Rather than requiring users to actively probe the environment, these techniques support users in building spatial awareness through continuous embodied interaction.

\subsection{Virtual World Setup}
\softwarename{}'s interactions are designed around three spatial concepts: (1) regions, (2) sections, and (3) objects. 

\subsubsection{\textbf{Region}}
A region represents a distinct, bounded area within an environment, capturing semantically meaningful spaces for users. For example, bedrooms, bathrooms, and hallways can each be considered regions in a house. Each region consists of two primary elements: sections and objects. Regions track which sections are explored by the user and keep the locations of discovered objects, enabling users to navigate to them. 
When a user enters a region for the first time, a TTS description provides a brief spatial overview of the region to support understanding of the region's layout and context (e.g., \emph{``You are in a large garden area surrounded by tall hedge walls, with a wooden table. The ground beneath you is a dirt path.''}). In Unity, a region is defined as an invisible 3D cube with an attached Unity script that tracks 
which objects and sections within it have been discovered. 

\subsubsection{\textbf{Section}} A section, defined as an invisible 3D cube in Unity, is a smaller spatial subdivision that organizes regions into discovered and undiscovered areas, helping users navigate to specific areas within a larger region.
While sighted users rely on visual architectural cues such as open corners or doorways to infer adjacent spaces, BLV users cannot. We address this by designing the sections to divide a region into smaller areas that have or have not yet been explored based on the user’s head movement. Specifically, in our design, 
sections can span multiple regions. By extending slightly beyond region boundaries, such as 
through doorways or corridors, sections signal the presence of nearby, unperceived space. When the headset detects that a section spans into a region other than the one the user currently occupies, the system indicates that additional areas are available to explore. Users can also navigate directly to undiscovered sections. 

\subsubsection{\textbf{Interactable Object}} An interactable object is any item in the virtual environment that users can engage with. Some objects are grabbable, allowing users to pick them up, while others are activated through controller or headset contact. Objects are located in sections and are discovered when explored by the user.

\subsection{Discovering the Environment} 
\label{sec:discoveryphase}
\softwarename{} enables users to reveal and catalog new regions and objects within the headset camera’s view by tuning their heads, dynamically updating the \emph{discovery menu} to guide further exploration. In this work, we conceptualize discovery as a self-directed process of acquiring spatial knowledge, in which users progressively uncover information about the environment through embodied actions, rather than being presented with all environmental information at once. Accordingly, we use head orientation as a proxy for users' current focus of attention to provide information in response to user exploration.

\subsubsection{\textbf{Camera-Based Discovery}}
The system identifies sections and objects based on the user’s head orientation by checking whether their bounds intersect the camera’s frustum and are not occluded by obstacles (e.g., walls). To accommodate diverse VR layouts, it does not impose a maximum detection distance; any unoccluded object within the camera frustum can be identified. When identified, the corresponding section and its objects are added to the \emph{discovery menu}. A chime notifies the user of a newly discovered section within a region, followed by audio cues announcing the names of any detected objects sequentially. Once a section is discovered, it is marked as such, and the system tracks remaining undiscovered sections that belong to that region. 

\subsubsection{\textbf{Dynamic Discovery Menu}} 
The \emph{discovery menu} serves as a memory aid for user exploration, listing all discovered regions, sections, and objects. Users can open it by pressing Y-button on the left controller to review unlocked (discovered) regions. When hovering over a region, the system provides a brief summary, including its name, whether it is fully discovered, and number of discovered objects; if a region has no discovered elements, it announces the region as empty. Selecting a region presents a prioritized list, with “Undiscovered Section” (if any remain) followed by discovered objects, all sorted by their proximity to the user. This structure encourages users to progressively explore hidden areas. The \emph{discovery menu} is dynamically updated as the user discovers more sections and navigates the environment. Specifically, as the camera detects new sections or objects, the system adds them to their parent region's list, and sorts the list by proximity to the user, with the closest items appearing first. The system also includes a Speech Rate Menu, allowing users to adjust TTS playback using five rates: 1x, 1.25x, 1.5x, 2x, and 4x. The selected rate applies globally and persists throughout the session, supporting individual preferences. 

\subsection{Navigating to Distant Targets} 
Users may either explore the environment freely or select a destination from the discovery menu for guided navigation. 
Once a destination is selected, \softwarename{} guides BLV users to distant sections and objects using automatically generated waypoints to define routes, along with responsive audio beacons that helps orient them toward the waypoints along the path. Rather than instantly transporting users to the selected destination, this design choice prioritizes active participation and a sense of control over minimizing navigation effort. Guiding users along the route also provides opportunities to discover nearby regions and objects that might otherwise be missed.

\subsubsection{\textbf{Waypoint-Based Navigation}}
When a distant target is selected from the \emph{discovery menu}, a navigation path is generated using Unity’s NavMesh~\cite{navmesh}, and waypoints are placed along this path. 
In pilot studies, initial paths frequently exhibited insufficient waypoint density around corners, leading to sharp turns that were difficult to navigate, and excessive waypoint density along straight segments, which hindered navigation by requiring users to stop and reorient after minor deviations.  
To address this, \softwarename{} refines waypoint placement through post-processing. The system first simplifies the navigation path by removing redundant waypoints that are closely spaced and follow nearly the same direction. It then improves the smoothness of movement by replacing sharp turns with gradual, curved transitions. To do this, it identifies turning points and adjusts the path by introducing additional waypoints that guide the user along a smoother trajectory. The system also refines waypoint placement relative to the user and the environment. Waypoints that are too close to the user are removed to avoid unnecessary guidance, while large gaps between waypoints are filled to maintain consistent navigation support. Together, these adjustments balance guidance and efficiency, making waypoint navigation more intuitive and reliable.

\softwarename{} spawns waypoints sequentially along the computed path. Before placing a waypoint, the system verifies that it aligns properly with the ground relative to the user. If both the user and the waypoint are on similar ground levels, the waypoint is positioned just below eye level. Otherwise, the system makes small vertical adjustments to ensure the waypoint rests above the ground and avoid collisions with overhead virtual obstacles. Waypoints are displayed as yellow spheres along the navigation path.
When the user reaches a waypoint, detected via headset or controller contact, it despawns, and a TTS cue confirms progress (e.g., \emph{“Reached Waypoint”}) along with the remaining waypoint count. This triggers the next waypoint to spawn and the cycle continues until all waypoints are cleared. 
Once the user reaches the destination, TTS announces the number of regions and objects discovered during navigation.

\subsubsection{\textbf{Orienting via Responsive Audio Beacon}}
\label{sec:sonification}
To help users orient toward waypoints, \softwarename{} extends traditional audio beacons with responsive, spatialized humming cues. Prior work uses audio beacons (sounds emitted from fixed 3D locations) to indicate object direction. In our system, each waypoint emits a continuous hum that dynamically adapts to the user’s head orientation. 
The hum varies in pitch and volume based on the angle between the user’s viewing direction and the waypoint. It reaches its highest pitch and volume when the user faces the waypoint and weakens as they turn away. We discretize this mapping into four tonal levels corresponding to increasing angular deviation (0–10°, 10–45°, 45–90°, and >90°), creating a clear directional gradient that complements spatial audio and supports efficient head alignment. 
When a user is far from a waypoint (>2 m), \softwarename{} takes the user's orientation to the waypoint to calculate angular deviations for playing the hums. At close range, vertical differences are ignored to prevent abrupt changes in sound (e.g., sudden decrease in hum when moving beneath a waypoint). At close distances, precise localization is supported through hand-based haptic cues.

For locomotion, users move forward using a joystick in the direction they are facing, as determined by their head orientation. We adopt joystick-based locomotion because it is simple, slightly preferred by BLV users~\cite{ribeiro2024investigating}, and well-suited to our virtual environment. Users orient to waypoints via responsive audio and move with the joystick. Each waypoint despawns on arrival, triggering the next until the target is reached. We play footstep sounds during locomotion to provide feedback on user movement and reinforce a sense of navigation.

\subsection{Localizing and Interacting with Nearby Items}
\subsubsection{\textbf{Localizing Nearby Items via Responsive Haptics}} 
\label{sec:controllers}
When the user is near a target (waypoint or object), controller vibration activates to support precise 3D localization alongside the spatial audio hum. We use vibration-based guidance for nearby items, as its physical cues align with how BLV users naturally locate nearby objects through touch in the real world. Users feel continuous vibration, with intensity primarily adapted by how accurately the controller is aimed toward the target. By moving the controller, BLV users can scan the space and use the responsive haptic cues to locate the target. 
In Unity, \softwarename{} computes the angular distance between each controller and the target. Once within 2 m, it calculates the angle between the controller’s forward vector (i.e., pointing direction) and the target vector (originating at the controller’s position and pointing toward the object). Haptic feedback is enabled when this angle is within ~90°. 
The vibration intensity is determined based on both the controller’s angular (85\% of intensity) and Euclidean (15\%) distance, providing a clear directional cue for fine-grained targeting. 
Together, audio provides directional guidance based on head orientation at distance (>2m), and haptics convey direction and proximity near the target.

\subsubsection{\textbf{Interacting with Objects and Obstacles}}
\label{sec:finalstage}
If the user selects an object as the navigation target, they can engage with it upon passing all waypoints and reaching the target. When a controller makes contact with the object, the responsive audio hum and haptic feedback stop, and TTS announces the object. For grabbable items, TTS indicates \emph{“Grabbable Object X,”} where X is the object’s name. Users can pick up the object using the grip button on either controller, with confirmation provided via \emph{“Object Grabbed X.”} Once grabbed, the object is added to the user’s inventory without needing them to hold the grip button longer. Other objects, such as gateways or consoles, are activated through controller or headset contact. 
After reaching the target, TTS also reports the number of regions and objects discovered during navigation.

To convey virtual walls and obstacles, the controllers provide feedback upon boundary contact. When a controller collides with a boundary, a short vibration pulse is triggered in that controller, accompanied by a spatialized “thud” sound. If the continuous vibration for object-localization is active, this guidance vibration pauses, a distinct collision pulse and thud are delivered, and the guidance vibration then resumes.

This combination of head-based environment discovery, waypoint-guided navigation, and responsive audio, haptic, and TTS feedback enables BLV users to independently explore and interact in VR.

\section{User Study}
We conducted a user study with \numparticipants{} BLV participants to investigate user experiences with \softwarename{}'s discovery and multimodal feedback mechanisms (RQ1) and the impact of free-form exploration on task performance and workload (RQ2). Sessions lasted 75 minutes, and participants received \$50. The study was approved by our institution’s IRB. 

\paragraph{\textbf{Participants}}
We recruited \numparticipants{} BLV participants ({5 male, 7 female}; ages 24-75) through a local blind community and service center (Table~\ref{tab:participants}). Eight participants identified as low vision and four as blind. All participants were over 18 and met the Social Security Administration’s criteria~\cite{ssa1980blindness} for legal blindness (central vision $\leq$ 20/200 or visual field $\leq$ 20°). All reported no impairments in touch perception. One participant (P1) had an auditory impairment but used a hearing aid; others reported no auditory impairments. 

\begin{table*}
\footnotesize
  \caption{Demographics of the \numparticipants{} participants in the user study. Participant numbers are sub-indexed by vision status: Blind (B), and Low Vision (LV).}
  \label{tab:Demographic Information}
  \centering
  \resizebox{\linewidth}{!}{\begin{tabular}
  {lllllll}
    \toprule
    \textbf{P\#} & \textbf{Age} & \textbf{Gender} & \textbf{Visual Impairment (VI)} & \textbf{VI Onset (Age)} & \textbf{VR Experience} & \textbf{Use of Digital Devices} \\
    \midrule
    P1\textsubscript{LV} & 65 & Male & Cataracts & 61 & No & Daily \\
    P2\textsubscript{LV} & 73 & Female & Glaucoma & 53 & No & Daily \\
    P3\textsubscript{B} & 62 & Female & Congenital Glaucoma & Birth & No & Several Times a Week \\
    P4\textsubscript{B} & 50 & Female & Blind with Residual Contrast Perception & 48 & Yes & Daily \\
    P5\textsubscript{LV} & 36 & Female & Uveitis, Glaucoma, and Cataracts & 26 & No & Daily \\
    P6\textsubscript{LV} & 45 & Male & Coloboma & Birth & No & Daily \\
    P7\textsubscript{LV} & 24 & Female & Aniridia and Nystagmus & Birth & Yes & Daily \\
    P8\textsubscript{LV} & 35 & Male & Optimistic Atrophy & 2 & No & Daily \\
    P9\textsubscript{B} & 25 & Male & ABCA4 Retinopathy & 4 & Yes & Daily \\
    P10\textsubscript{LV} & 75 & Female & Age-Related Macular Degeneration & 72 & Yes & Daily \\
    P11\textsubscript{LV} & 32 & Female & Stickler Syndrome & Birth & No & Daily \\
    P12\textsubscript{B} & 47 & Male & Total Blindness & Birth & No & Daily \\

    \bottomrule
  \end{tabular}}%
  \label{tab:participants}
\end{table*}

\paragraph{\textbf{Study Design}}
We used a within-subject design to evaluate the two primary design components of VisionPulse: exploration type and feedback modality. 
For exploration type, we compared (1) \softwarename's \emph{discovery}, where participants uncovered regions and objects through exploration, which were progressively added to the menu (Section~\ref{sec:discoveryphase}) against (2) \emph{prebuilt menu}, where all regions and objects were available in the navigation menu from the start, serving as a baseline based on prior work. 
For feedback modality, we compared: (1) \emph{audio beacon}, using standard 3D spatialized audio from the target as a baseline based on prior work; (2) \emph{responsive audio beacon}, which dynamically adjusts pitch and volume based on head orientation (Section~\ref{sec:sonification}); and (3) \softwarename{}'s \emph{responsive audio beacon with haptics}, which adds controller-based vibrations (Section~\ref{sec:controllers}). We included the responsive audio beacon condition without haptics to isolate the contribution of haptic feedback and assess its impact on BLV users' experience.
Together, these factors formed six experimental conditions. 

\begin{figure*}[htbp]
    \centering

    \begin{subfigure}{0.32\textwidth}
        \centering
        \includegraphics[width=\linewidth]{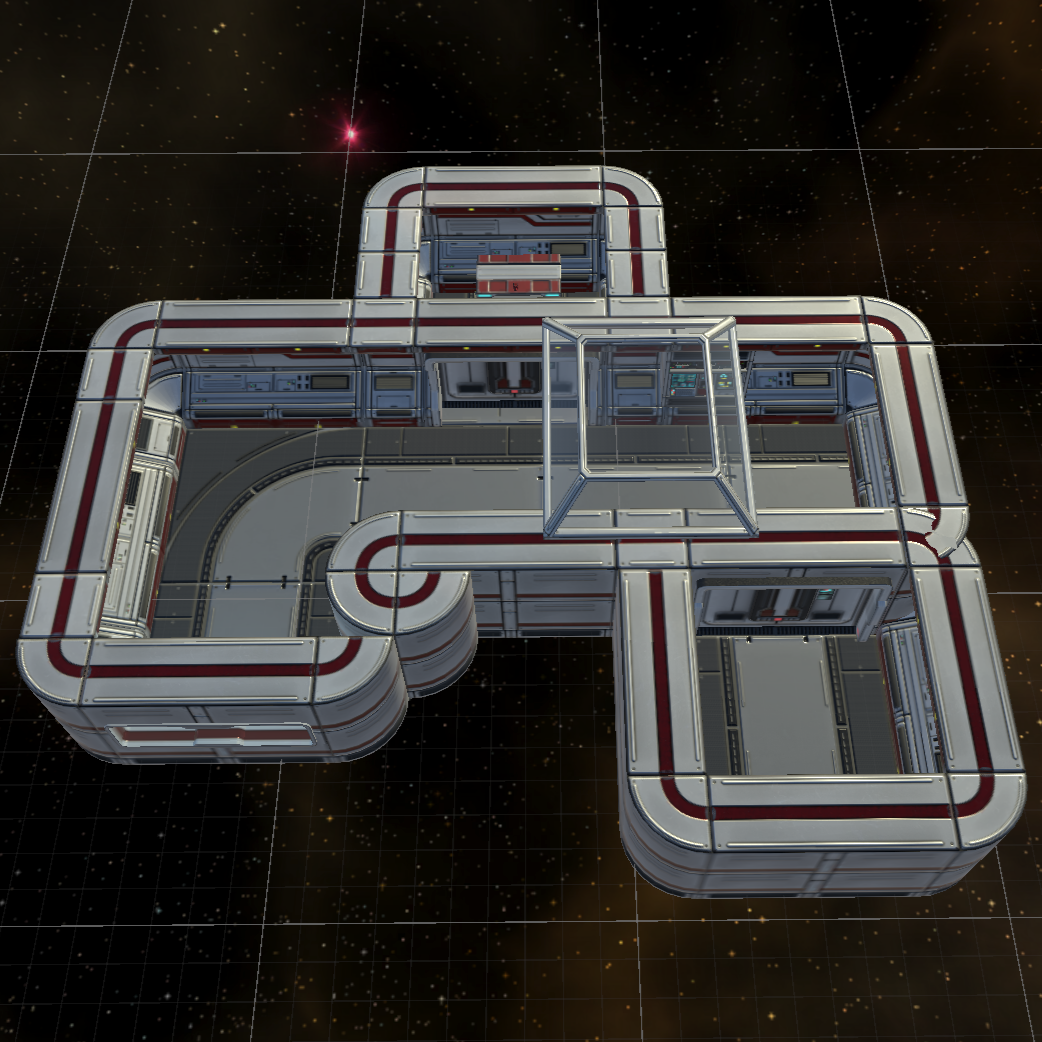}
        \label{fig:sub1}
    \end{subfigure}
    \hfill
    \begin{subfigure}{0.32\textwidth}
        \centering
        \includegraphics[width=\linewidth]{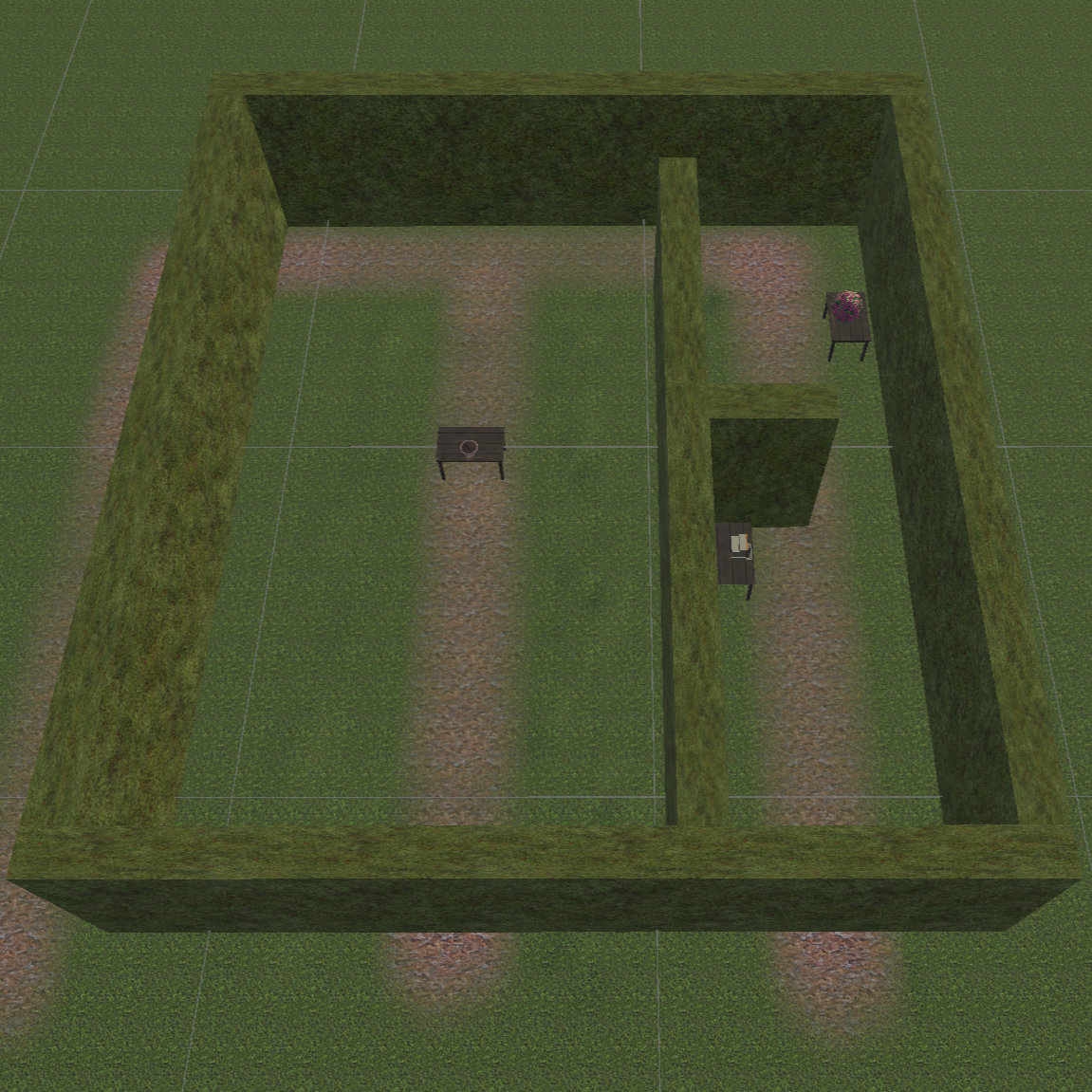}
        \label{fig:sub2}
    \end{subfigure}
    \hfill
    \begin{subfigure}{0.32\textwidth}
        \centering
        \includegraphics[width=\linewidth]{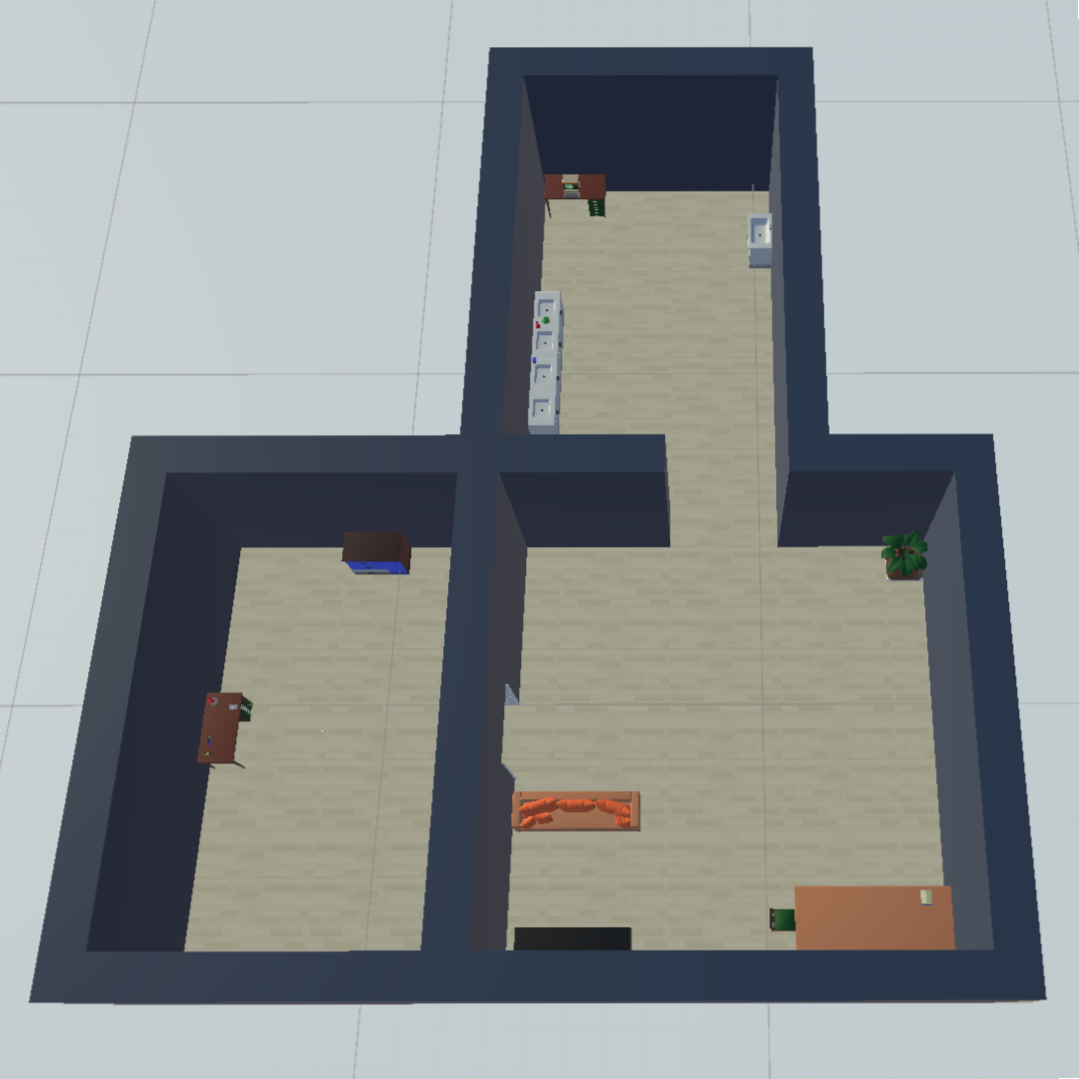}
        \label{fig:sub3}
    \end{subfigure}

     \begin{subfigure}{0.32\textwidth}
        \centering
        \includegraphics[width=\linewidth]{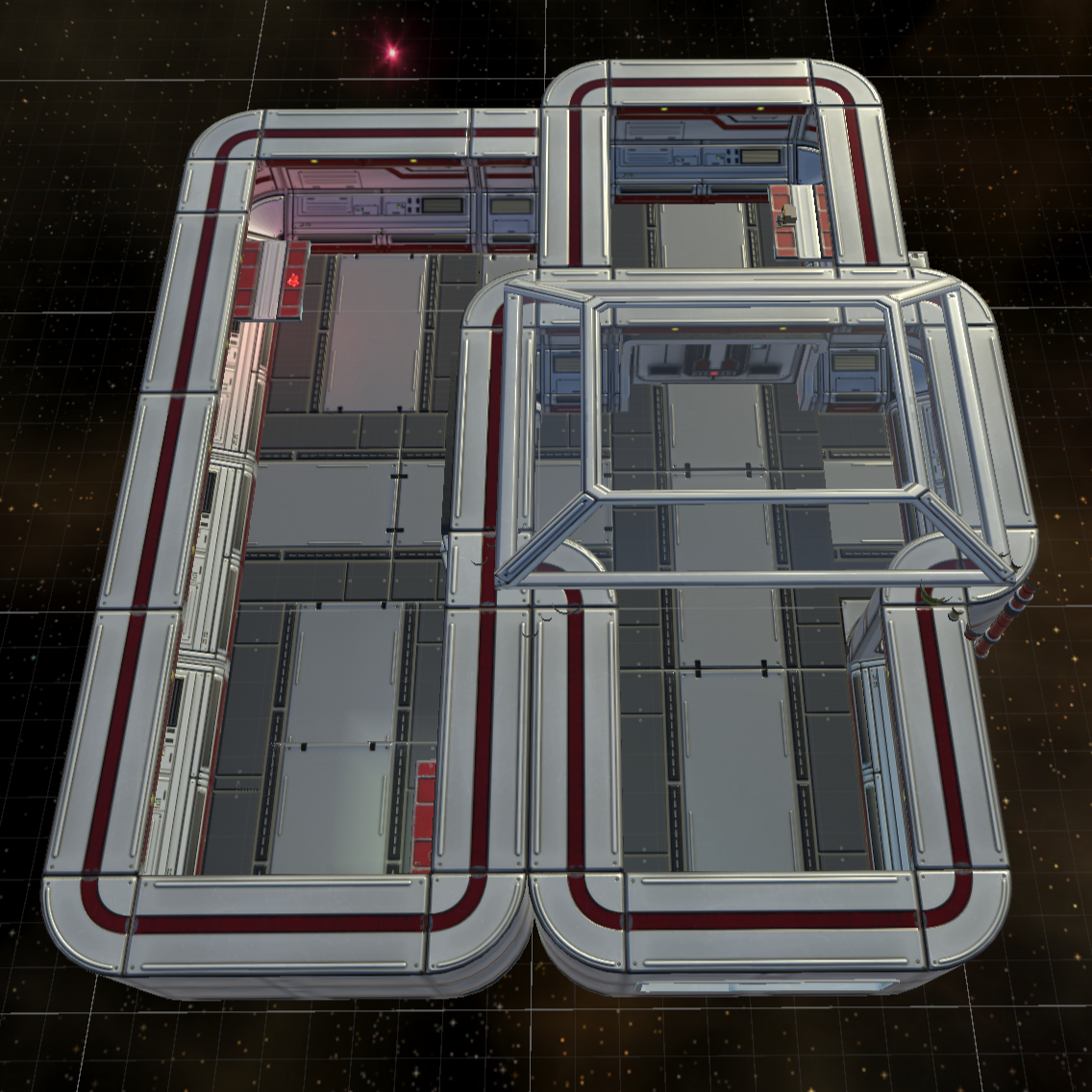}
        \label{fig:sub4}
    \end{subfigure}
    \hfill
    \begin{subfigure}{0.32\textwidth}
        \centering
        \includegraphics[width=\linewidth]{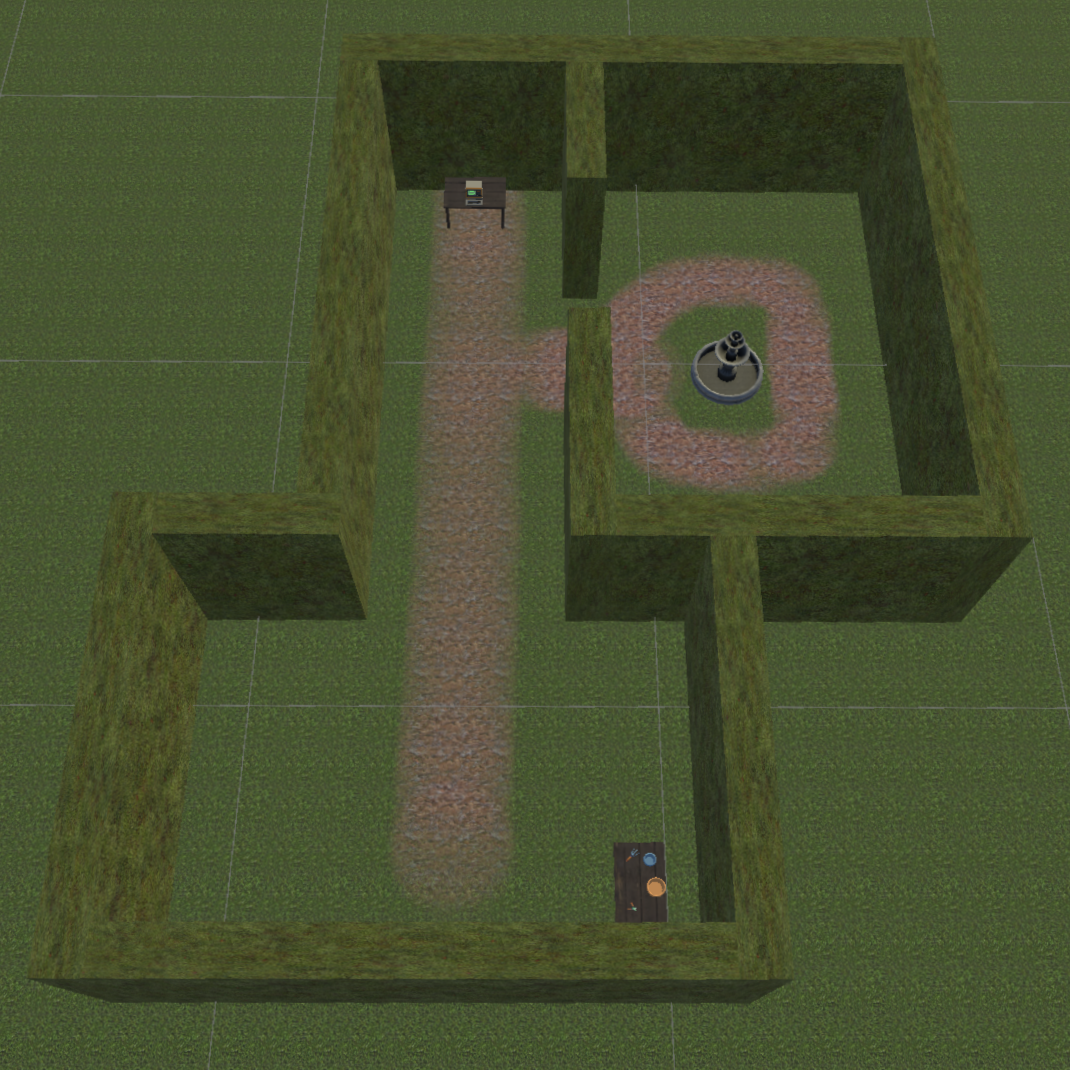}
        \label{fig:sub5}
    \end{subfigure}
    \hfill
    \begin{subfigure}{0.32\textwidth}
        \centering
        \includegraphics[width=\linewidth]{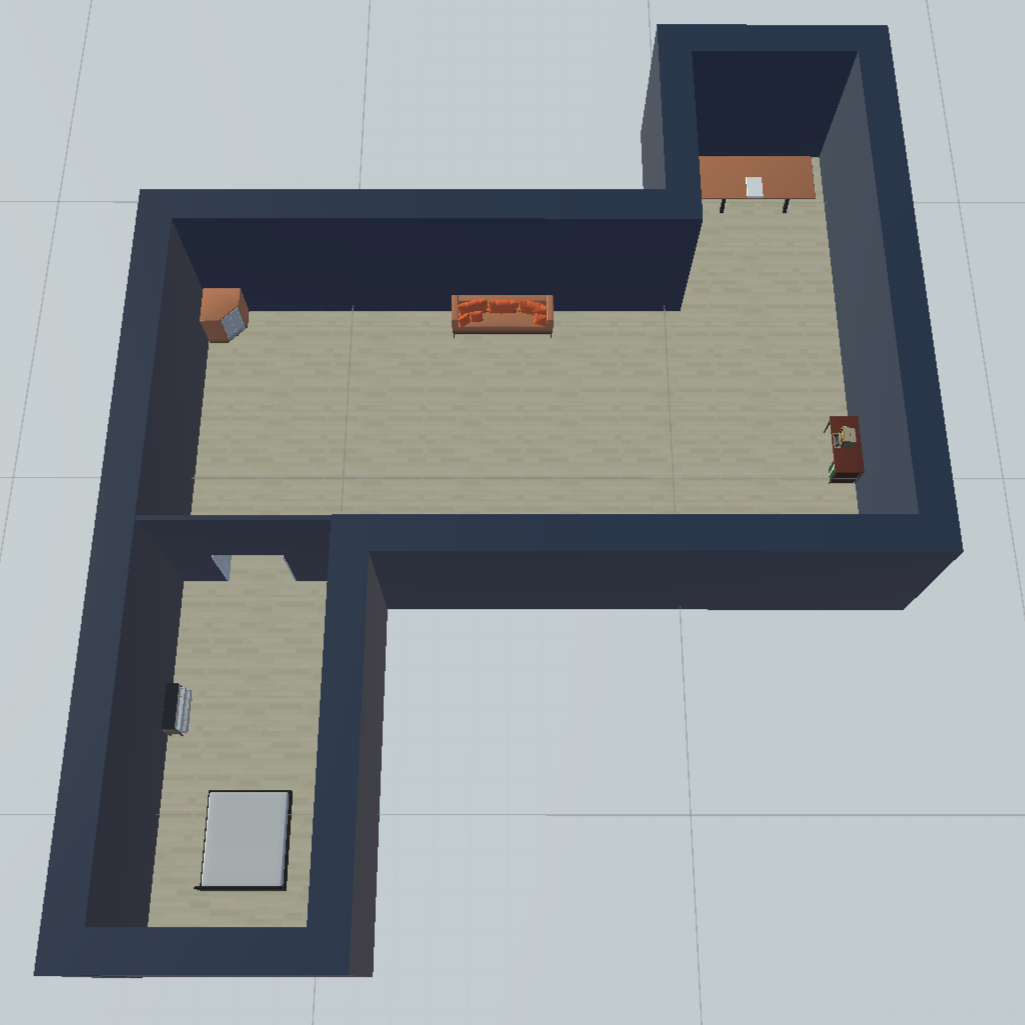}
        \label{fig:sub6}
    \end{subfigure}

     \caption{Six VR scenes with distinct themes: two spaceships, two gardens, and two houses. We varied themes to prevent boredom while keeping scene specifications consistent.}
    \label{fig:vrscenes}
\end{figure*}

\paragraph{\textbf{VR Scenes and Task}}
We designed six VR scenes with distinct themes: two gardens, two spaceships, and two houses (Figure~\ref{fig:vrscenes}). All scenes followed consistent structural and spatial properties to ensure comparable navigation scale and complexity. Each scene included three regions and three primary objects (a key, a gateway, and a console), with 2-3 of these objects not in the participant's view when they start the scene. Scene perimeters ranged from 80–95 m, volumes from 1300–1800 m$^3$, and average path lengths to targets from 15–25 m. 

In each scene, the study task required locating a key object and then navigating to a gateway object within 10 minutes. Attempting to interact with the gateway object without collecting the key triggered a TTS warning \emph{“The key has not been collected.”} Successful completion announced, \emph{``The key has been collected. Moving on to the next scene.''} Scene order was randomized. 
Interacting with the console was optional and not required to complete the scene. Each console presented a brief, theme-related fun fact (e.g., \emph{``The Gardens of Versailles in France are the largest garden in the world, covering more than 800 hectares of land.''}). We included the console to assess if different conditions influenced user willingness to engage with non-essential elements. 

These design choices ensured consistent task difficulty, navigation demands, and spatial scale across conditions while allowing for thematic variation.

\begin{figure}[htbp]
    \centering

    \begin{subfigure}[t]{0.46\linewidth}
        \centering
        \includegraphics[width=\linewidth]{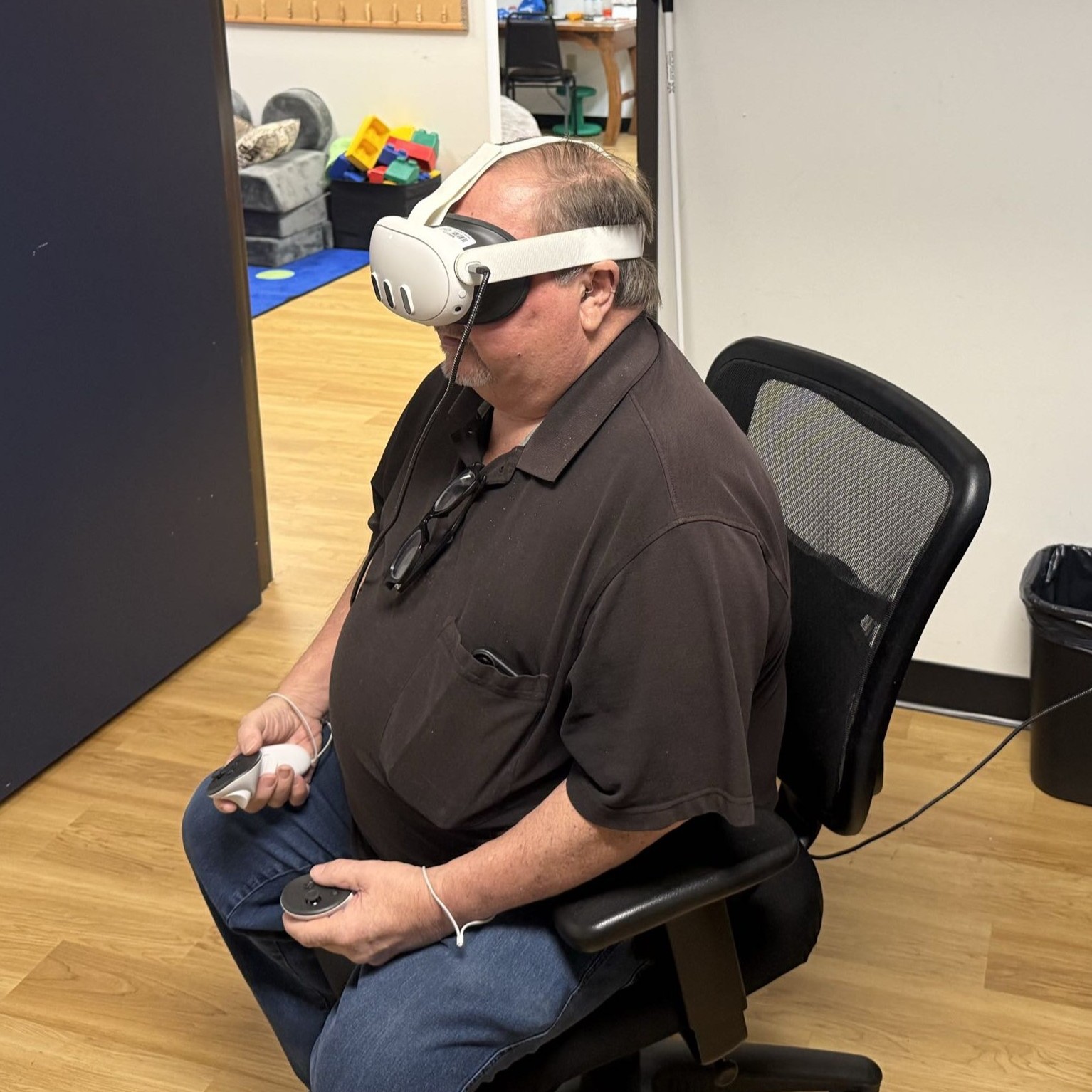}
    \end{subfigure}
    \hspace{0.02\linewidth}
    \begin{subfigure}[t]{0.46\linewidth}
        \centering
        \includegraphics[width=\linewidth]{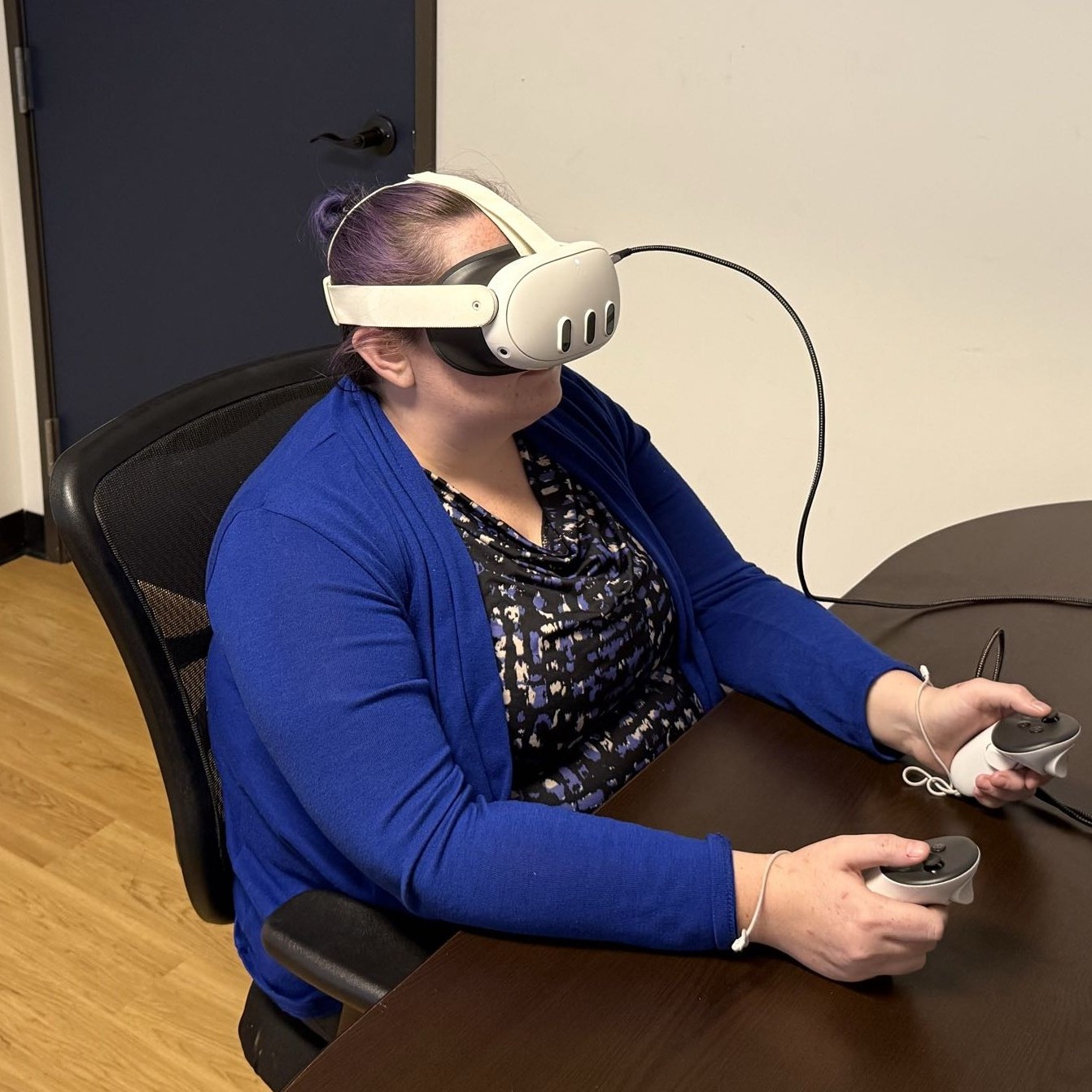}
    \end{subfigure}

    \caption{User study setup. Participants wore the headset and explored the VR environment in a seated position. They oriented themselves by rotating the chair and navigated using joystick-based locomotion.} 
    \label{fig:participantsusingvr}
\end{figure}

\paragraph{\textbf{Procedure}}
\label{sec:procedure}
The study consisted of three stages: (1) introduction with a pre-questionnaire and tutorial, (2) the main experiment with six conditions, and (3) a post-study questionnaire and interview. Participants were seated in a swivel chair, allowing them to physically rotate for exploring their surrounding but kept them stationary to minimize collision risk and balance issues.

After informed consent, participants completed a pre-questionnaire on their demographics, vision status, and VR experience, followed by an introduction to \softwarename{}. They practiced in a tutorial scene, including a small white room containing a red cube and a green cube. They were instructed to locate and grab the red cube and then navigate to the green cube to complete the tutorial. 
This scene enabled them to complete warm-up tasks such as menu use, target selection, navigation, and object interaction in a simplified environment.
In the main experiment, participants completed the VR task under one exploration type (\emph{prebuilt} or \emph{discovery}) in three scenes, each with a different feedback modality. Before each modality, they practiced the feedback in the tutorial scene before moving to the experimental scene. They then repeated the process with the alternate exploration type. For each scene, if participants exceeded the 10-minute limit or chose to stop due to frustration, they were advanced to the next scene. We counterbalanced the condition order (exploration type, feedback modality) and randomized the assignment of conditions to VR scenes.
After each scene, participants completed two questionnaires on spatial presence (MEC-SPQ) and workload (NASA TLX). They also rated their perceived agency and autonomy after each exploration type. Finally, they participated in a semi-structured interview with open-ended questions about their preferences between the two exploration conditions and feedback modalities, the reasons for those preferences, the aspects of the system they found most helpful or challenging, suggested improvements, and their overall experience.

\paragraph{\textbf{Data Collection and Analysis}}
We collected task data, subjective ratings, and interview responses. Task data included participants’ positional logs (for path length), completion time, and console interactions. Task completion time was defined as the duration from scene start to successful gateway interaction after key collection. We also collected subjective ratings using a modified MEC-SPQ for spatial presence~\cite{vorderer2004mec} and NASA-TLX for workload~\cite{peter1988development} questionnaires. We additionally collected user agency and autonomy ratings on the following statements using 1 to 5 Likert scales: (1) I felt in control of my actions in the virtual environment; (2) I enjoyed navigating and interacting with the virtual environment; and (3) I felt a sense of accomplishment when discovering the virtual environment. Interviews were recorded and transcribed. 

We analyzed navigation patterns and interview data qualitatively. Using participants' positional data, we categorized their movement into three types: (a) \emph{deliberate} includes goal-directed navigation to a target with minimal deviation; (b) \emph{wandering} includes navigation to a target with a few turns, detours, and occasional revisiting of previously traversed areas; and (c) \emph{erratic} captures highly irregular and unpredictable trajectories, including sharp turns, zig-zagging, and substantial overlap with previously traveled space, indicating little consistent progression toward a goal. 
Two authors independently coded movement patterns for all study sessions, then resolved discrepancies through discussion, and consolidated the labels. 
Interview data were analyzed using thematic analysis inspired by Braun and Clarke's approach~\cite{clarke2021thematic}. We applied open coding to all transcripts using MAXQDA qualitative analysis software. After the initial coding, two authors reviewed the codes, discussed any differences and patterns in the data, then one author drafted the themes. These themes were subsequently reviewed and refined through iterative reflection with the whole team, using the coded data for reference.

The collected measures helped address our two research questions. To answer RQ1, we analyzed participants' preferences, spatial presence, agency, and interview responses to understand how BLV users experienced \softwarename{}'s discovery and multimodal feedback. To answer RQ2, we analyzed task completion time, path length, movement patterns, and perceived workload to evaluate how discovery-based exploration affected task performance and workload compared to a pre-discovered environment.

\section{Results}
We present the quantitative results followed by the qualitative themes derived from the interviews. Throughout this section, participant identifiers are sub-indexed by vision status: Blind (B), and Low Vision (LV).

\subsection{Quantitative Results}

\begin{table*}[h]
\centering
\caption{Statistical analyses on the effects of Exploration Type and Feedback Modality on task performance and subjective measures. We used a two-way repeated-measures ANOVA for NASA-TLX and ART ANOVAs for all other measures due to non-normality.}
\label{tab:ANOVA}
\begin{tabular}{llccccc}
\toprule
\textbf{Factors} & \textbf{Measures} & $df_1$ & $df_2$ & $F$ & $p$ & \textbf{\textbf{$\eta_{p}^{2}$}} \\
\toprule
\multirow{5}{*}{Exploration Type}     
                        & Completion Time & 1 & 48 & 1.73 & .19 & .310 \\
                        & Path Length & \textbf{1} & \textbf{55} & \textbf{9.53} & \textbf{.003} & \textbf{.150} \\
                        & NASA-TLX & 1 & 11 & .500 & .492 & .044 \\
                        & MEC-SPQ (SSM) & 1 & 55 & .830 & .37 & .010 \\
                        & MEC-SPQ (SPSL) & 1 & 55 & .270 & .61 & $<$.010 \\
                        \cmidrule(lr){1-7}
\multirow{5}{*}{Feedback Modality}     
                        & Completion Time & 2 & 48 & .950 & .390 & .040 \\
                        & Path Length & 2 & 55 & 3.050 & .056 & .100 \\
                        & NASA-TLX & 2 & 22 & .250 & .782 & .022 \\
                        & MEC-SPQ (SSM) & 2 & 55 & .750 & .430 & .030 \\
                        & MEC-SPQ (SPSL) & 2 & 55 & .066 & .940 & $<$.010 \\
                        \cmidrule(lr){1-7}
\multirow{5}{*}{Exploration Type*Feedback Modality}  
                        & Completion Time & 2 & 48 & 1.940 & .150 & .070 \\
                        & Path Length & 2 & 55 & .320 & .729 & .010 \\
                        & NASA-TLX & 2 & 22 & .030 & .973 & .003 \\
                        & MEC-SPQ (SSM) & 2 & 55 & .580 & .560 & .020 \\
                        & MEC-SPQ (SPSL) & 2 & 55 & .950 & .390 & .030 \\
\bottomrule
\end{tabular}
\end{table*}

\paragraph{\textbf{Participants' Preferences}} Participants overwhelmingly preferred the \emph{discovery} exploration type, with 10 out of 12 selecting it over the \emph{prebuilt menu} (2/12). Similarly, 10 participants preferred the combination of \emph{responsive audio beacon with haptics}, while two preferred the \emph{responsive audio beacon alone}.

\paragraph{\textbf{Task Performance}} Table~\ref{tab:ANOVA} summarizes the statistical significance results. Task-related dependent variables violated normality assumptions (Shapiro–Wilk, $p$ $<$ .05), thus we used Aligned Rank Transform (ART) ANOVA \cite{wobbrock2011aligned} for non-parametric analysis. 
For path length, there was a significant main effect of exploration type, with longer paths observed in the \emph{discovery} ($M = 148.70m\pm100.89$) compared to the \emph{prebuilt} condition ($M = 97.39m\pm52.99$). The effect of feedback modality was marginally non-significant, suggesting similar navigation efficiency across modalities (\emph{audio beacon}: $M = 138.09m\pm80.60$; \emph{responsive audio beacon}: $M = 114.58m\pm96.32$; \emph{responsive audio beacon with haptics}: $M = 116.45m\pm73.08$). For completion time, we excluded seven experimental scenes in which participants exceeded the 10-minute limit or requested to skip due to frustration. The excluded scenes came from four participants (two blind and two low vision), all aged 50 or older. For three participants, the excluded scene occurred in the first experimental scene. Specifically, one participant (P3\textsubscript{B}) had four excluded scenes (two \emph{prebuilt} and two \emph{discovery} across various feedback conditions), while three participants (P4\textsubscript{B}, P2\textsubscript{LV}, and P10\textsubscript{LV}) each had one excluded scene in the \emph{discovery} condition with \emph{audio beacon} feedback. These unsuccessful trials likely reflected individual differences in familiarity with VR technology, the exploration tasks, or interpreting the multimodal feedback, making some scenes more challenging to complete within the allotted time. On average, participants completed the VR task in $M = 261.86s\pm148.16$ for \emph{prebuilt} and $M = 301.63s\pm184.07$ for \emph{discovery}. Across feedback modalities, task completion time was $M = 234.20s\pm123.93$ for \emph{audio beacon}, $M = 305.73s\pm197.01$ for \emph{responsive audio beacon}, and  $M = 305.30s\pm165.45$ for \emph{responsive audio beacon with haptics}. The differences were not significant. No other significant main effects or interaction effects were found, suggesting similar task performance regardless of experimental condition. Participant interaction with the console was similar across conditions (\emph{prebuilt}: 14 out of 36, \emph{discovery}: 13/36). On average, low vision users completed tasks faster than blind users (246 vs. 310 s) and explored more of the environment (119 vs. 88.5 m). We did not perform statistical comparisons between groups due to small, unequal samples. 

\paragraph{\textbf{Perceived Workload, Spatial Presence, and Agency}} NASA-TLX scores (0–100 scale, 5-point increments) satisfied the normality assumption. A two-way repeated-measures ANOVA revealed no significant main effects of exploration type, feedback modality, nor an interaction effect, indicating similar perceived workload across conditions (Table~\ref{tab:ANOVA}). Average NASA TLX scores were $25.79\pm 29.12$ for \emph{prebuilt} and $23.70\pm 26.63$ for \emph{discovery}, suggesting moderate workload. Spatial presence (MEC-SPQ) measures, namely Spatial Situation Model (SSM), which reflects the extent to which participants form a mental representation of the spatial layout of the mediated environment, and Spatial Presence Self Location (SPSL), which captures the degree to which participants feel physically located within the virtual environment, did not meet normality assumptions when we analyzed the data using an ART ANOVA. No significant effects were observed for SSM (\emph{prebuilt}: $M = 4.17\pm0.94$, \emph{discovery}: $M=4.03\pm1.03$). Similarly no effects were observed for SPSL factors (\emph{prebuilt}: $M = 4.19\pm0.95$, \emph{discovery}: $M = 4.24\pm0.97$). Participants reported high agency ratings across conditions (\emph{prebuilt}: $M = 4.58\pm.68$, \emph{discovery}: $M = 4.67\pm.58$). Workload and presence ratings were similar between blind and low vision users (NASA-TLX: 23.5 vs. 19.9; SSM: 4.1 vs. 4.2; SPSL: 4.3 vs. 4.1). 

\begin{table*}[h]
\centering
\caption{Distribution of movement behavior (\emph{deliberate}, \emph{wandering}, or \emph{erratic}) across exploration types and feedback modalities.} 
\label{tab:movement}
\begin{tabular}{llccc}
\toprule
\textbf{Exploration Type} & \textbf{Feedback Modality} & Deliberate & Wandering & Erratic \\
\toprule
\multirow{3}{*}{Prebuilt}     
                        & Audio beacon & 7 & 3 & 2 \\
                        & Responsive audio beacon & 11 & 1 & 0 \\
                        & Responsive audio beacon with haptics & 9 & 3 & 0 \\
                        \cmidrule(lr){1-5}
\multirow{3}{*}{Discovery}     
                        & Audio beacon & 8 & 1 & 3 \\
                        & Responsive audio beacon & 7 & 3 & 2 \\
                        & Responsive audio beacon with haptics & 5 & 5 & 2 \\
\bottomrule
\end{tabular}
\end{table*}

\begin{figure*}[htbp]
    \centering

    \begin{subfigure}{0.32\textwidth}
        \centering
        \includegraphics[width=\linewidth]{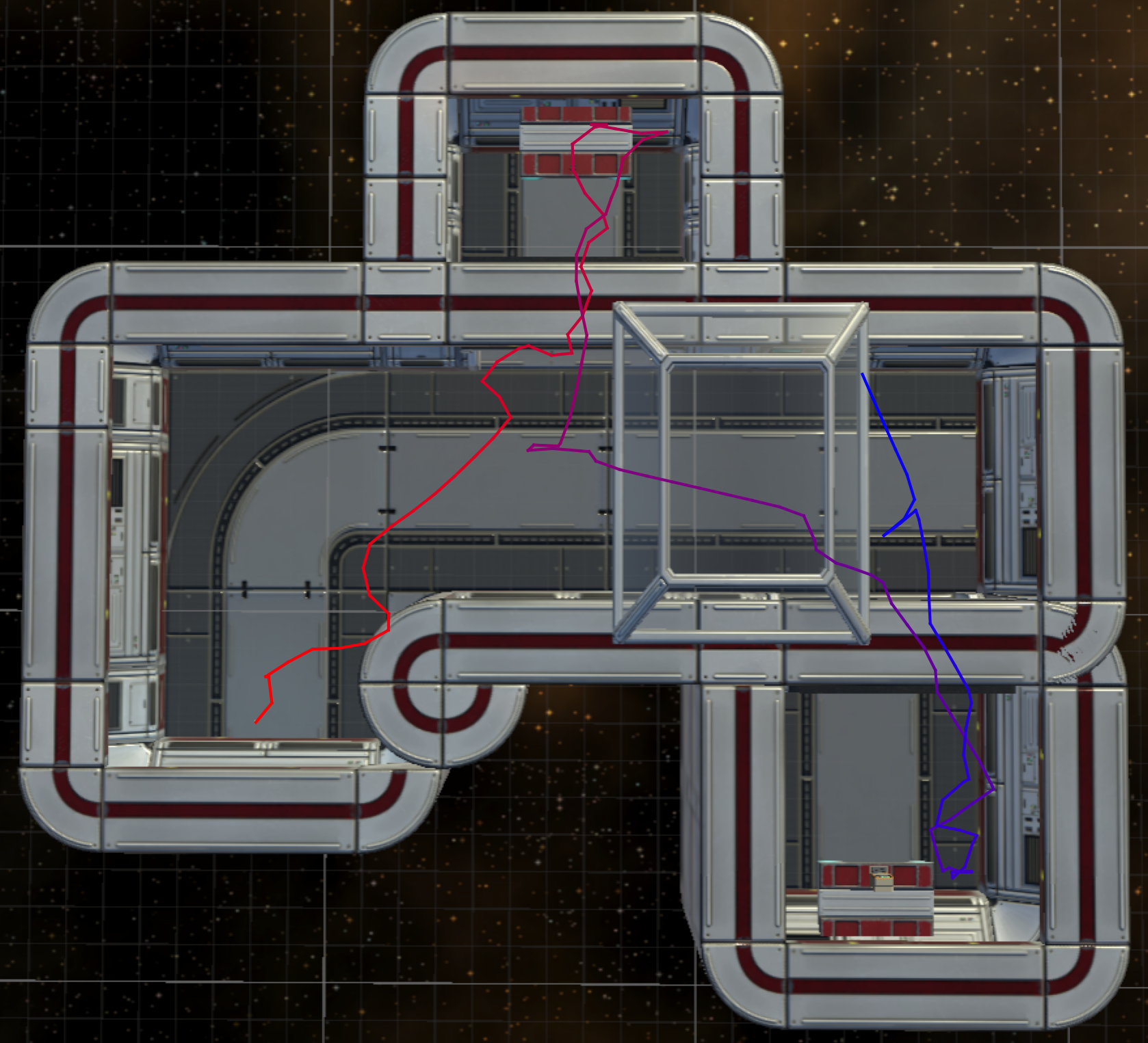}
        \caption{Deliberate}
        \label{fig:sub1}
    \end{subfigure}
    \hfill
    \begin{subfigure}{0.32\textwidth}
        \centering
        \includegraphics[width=\linewidth]{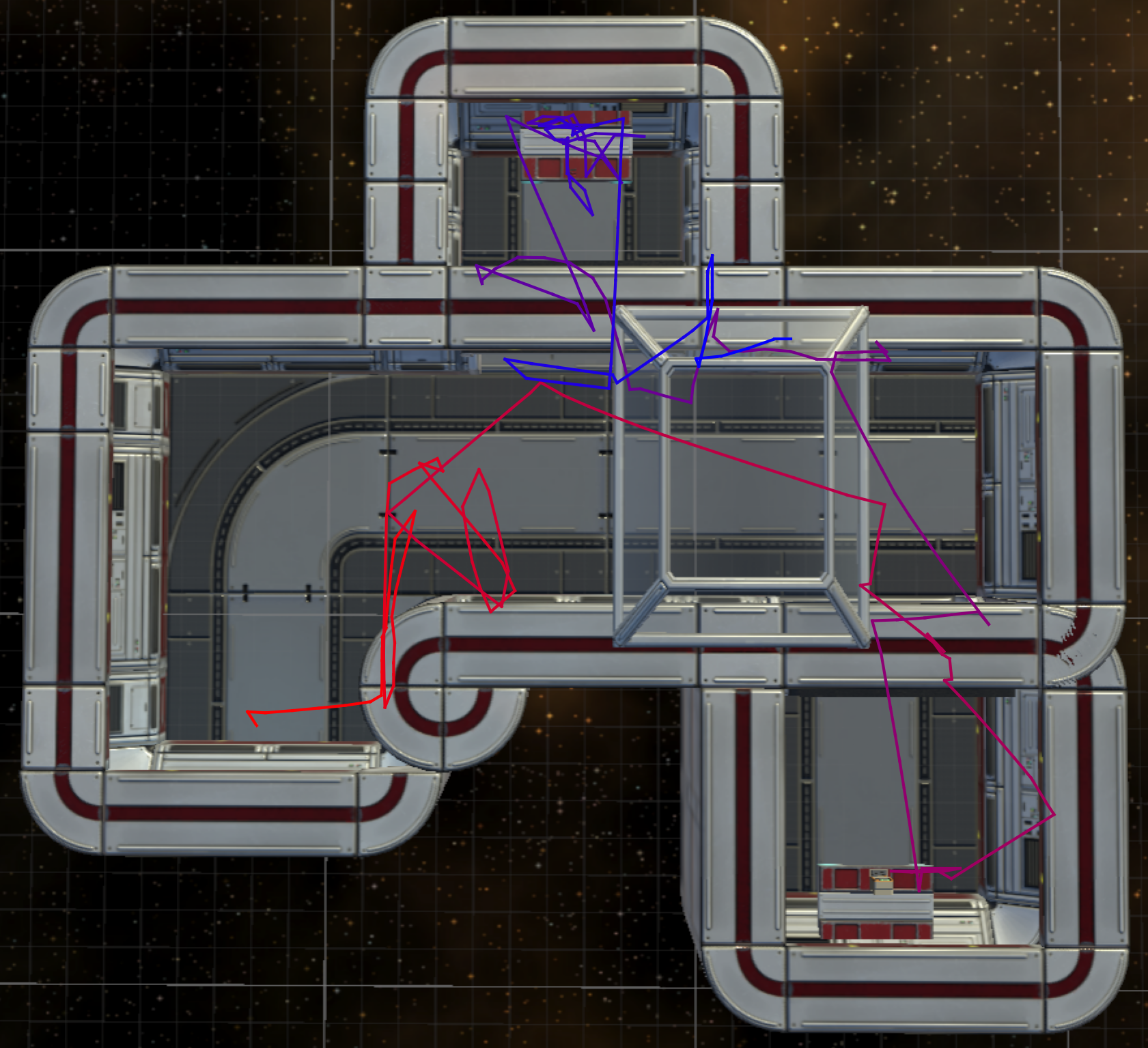}
        \caption{Wandering}
        \label{fig:sub2}
    \end{subfigure}
    \hfill
    \begin{subfigure}{0.32\textwidth}
        \centering
        \includegraphics[width=\linewidth]{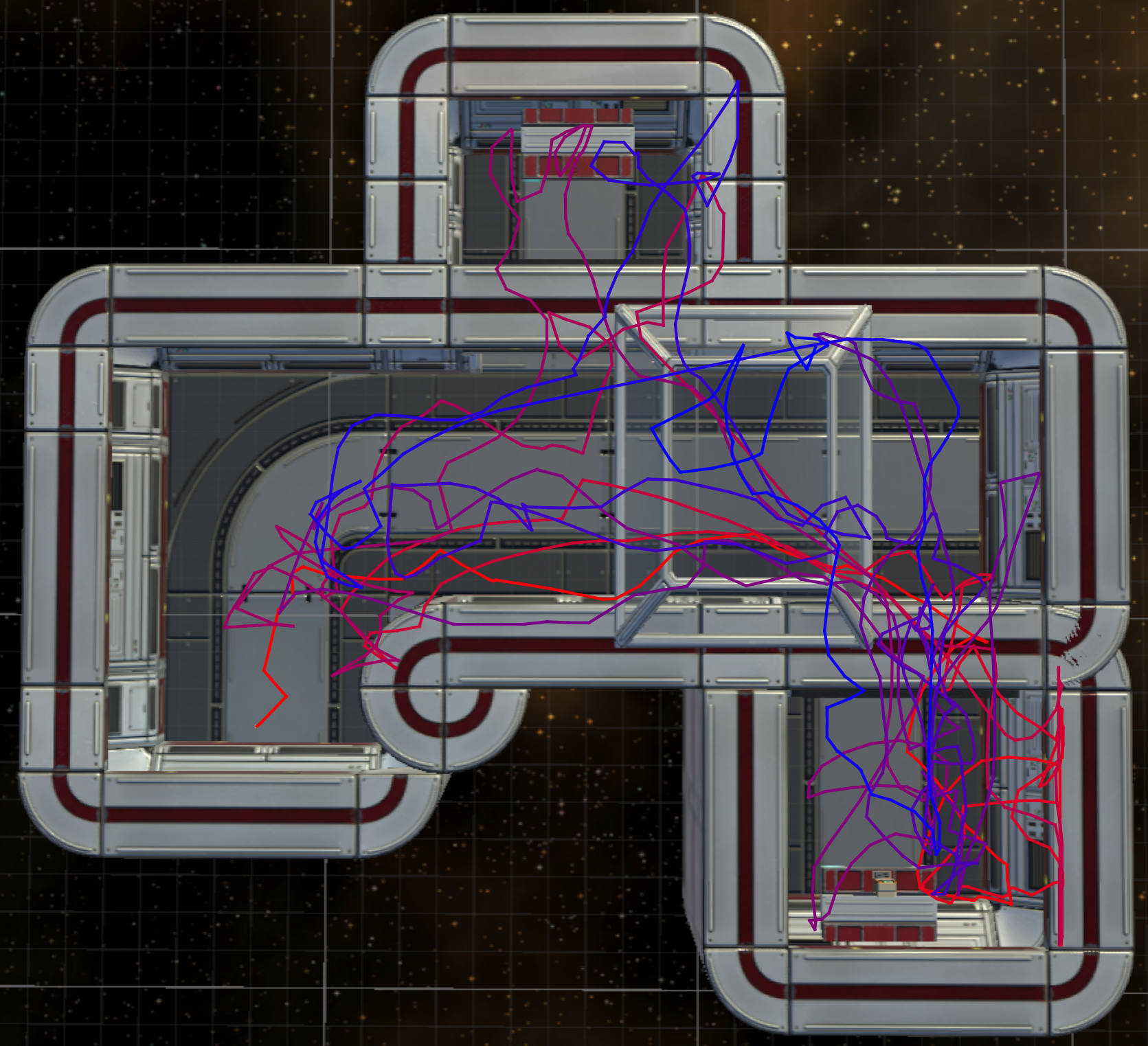}
        \caption{Erratic}
        \label{fig:sub3}
    \end{subfigure}

     \caption{Overview of movement patterns in a spaceship-themed scene. User paths are color-coded over time, progressing from red to purple to blue. Fig.~\ref{fig:movementtypes}\subref{fig:sub1} illustrates \emph{deliberate} movement, with a clear, goal-directed path from the start to the key, console, and gateway. Fig.~\ref{fig:movementtypes}\subref{fig:sub2} shows \emph{wandering} movement, where goal-directed progress is interspersed with zig-zagging and occasional loops. Fig.~\ref{fig:movementtypes}\subref{fig:sub3} depicts \emph{erratic} movement, characterized by highly unpredictable trajectories with no consistent pattern.} 
    \label{fig:movementtypes}
\end{figure*}

\paragraph{\textbf{Movement Patterns}} 
Table~\ref{tab:movement} summarizes participants’ movement patterns (Fig.~\ref{fig:movementtypes}) by exploration type and feedback modality. In the \emph{prebuilt} navigation conditions, \emph{deliberate} navigation was predominant across all modalities, with less \emph{wandering} and rare \emph{erratic} behavior (observed only with \emph{audio beacon}). In the \emph{discovery} conditions, \emph{deliberate} paths were again the most frequent, though movement types were more evenly distributed, suggesting that while \emph{prebuilt menu} conditions promote goal-directed navigation, \emph{discovery} can encourage broader exploration.

Across movement types, \emph{deliberate} paths ($n=47$) were associated with lower workload but also slightly lower spatial presence. Specifically, \emph{deliberate} paths had mean scores of NASA-TLX $= 23.17\pm25.85$, SSM $= 4.05\pm0.98$, SPSL $= 4.09\pm0.98$. In contrast, the \emph{wandering} pattern ($n = 16$) yielded higher perceived workload (NASA-TLX $= 27.66\pm30.23$), but also slightly higher spatial presence (SSM $= 4.16\pm0.97$; SPSL $= 4.47\pm0.79$). Similarly, after the \emph{erratic} pattern ($n = 9$), participants reported a higher workload (NASA-TLX $= 27.78\pm32.95$), but also higher spatial presence SSM $= 4.28\pm1.04$ and SPSL $= 4.44\pm1.01$. This suggests more exploration may contribute to a higher sense of spatial presence.
Regarding vision status, we did not observe meaningful differences in movement patterns between blind and low vision participants (B: 4.00 vs. LV: 3.88 \emph{deliberate} paths, on average). However, some participants with low vision reported using residual vision to identify scene elements, such as the yellow waypoint spheres, which reinforced the audio-haptic feedback.

\subsection{Qualitative Findings}
Participants consistently described their VR interactions as enjoyable and smooth, highlighting both usability and intrinsic engagement (n=11). Many emphasized the system’s reliability and low friction as key contributors to their positive experience, which fostered a strong desire for continued use. This was reflected in comments such as P2\textsubscript{LV}: “\textit{I wanted to keep doing it.}” 
Such enthusiasm was often expressed spontaneously, indicating excitement about \softwarename{}'s potential. For example, when describing their experience of the discovery condition, P12\textsubscript{B} noted, “\textit{I did not speed up the TTS, although I know I could have because I was so excited for what's coming up}”. Beyond enjoyment, participants also articulated broader value in the system. 
P8\textsubscript{LV} highlighted \softwarename{}'s potential for social inclusion: “\textit{I mentioned my buddy earlier. He’s really into gaming and stuff like that... Right now, what he does is say, ‘Hey, hop on Discord and just hang out with us,’ and I do... but I don’t feel included, because they’re playing and I’m just listening and watching. Whereas now, with this... I can say, ‘Hey, yeah—jump into VR. I’ll come play with you guys… and for once I won’t feel left out.}”

Below, we present three themes derived from interviews and observations, highlighting how exploration and navigation techniques, along with multimodal feedback, shape BLV users’ VR experiences, including their sense of autonomy, spatial awareness, and interaction.

\subsubsection{\textbf{Theme 1: Supporting User Autonomy by Combining Self-Directed Discovery with Guided Navigation}}
Most participants described discovery as a primary driver of engagement, with the process of uncovering the environment providing intrinsic satisfaction. Rather than relying on explicit instruction, participants valued figuring things out independently, often framing discovery as more meaningful than being told what to do. As P12\textsubscript{B} explained, “\textit{... discovery for me… I think that's the best way for people to really get their environment... I mean, it's okay to be told where things are, but sometimes it's just better to discover it and learn where it is.}” This preference aligned with real-world behaviors; as P11 noted, “\textit{ I like to explore my surroundings in real life, so in virtual worlds, I like to do the same.}”

A key aspect of this engagement was a strong preference for autonomy. Participants valued self-directed interaction and the ability to decide what to explore, which contributed to a sense of ownership and personal investment. As P11\textsubscript{LV} stated, “\textit{I like to make my own decisions and find things on my own and see what's important and not.}” Participants also described \softwarename{}’s navigation support as enabling rather than constraining independence. Participants described guidance features as supporting their intentions and helping them reach goals more confidently. For example, P4\textsubscript{B} noted, “\textit{I like the being able to pull up the menus and then being able to zone in on an object that I'm looking for once it was discovered... And then being able to help me get to that area.}” These findings suggest that autonomy in immersive environments emerges from the ability to combine self-directed exploration with supportive navigation tools.

Through exploration, participants described gradually constructing mental representations of the environment. Discovery-based interaction encouraged users to build spatial understanding through movement and feedback rather than predefined instructions. This process of navigating uncertainty fostered deeper cognitive engagement and supported mental map formation. As P12\textsubscript{B} explained, “\textit{ I liked kind of not exactly knowing where everything was, and I had to kind of learn where it was.}” Similarly, P6\textsubscript{LV} noted, “\textit{it [discovery] also kind of helps with mental mapping as well}”.

Some participants described discovery as challenging, but this challenge often enhanced engagement by introducing a game-like quality. P8\textsubscript{LV} remarked, “\textit{I felt like it was like an actual video game that I've seen. So you want to discover what abilities, what you can do in that world or in that game.}” Rather than viewing difficulty as a barrier, participants appreciated the need for active problem solving. P2 explained, “\textit{Because I had to think it out…more challenging.}” P4\textsubscript{B} added, “\textit{I like the mental challenge of discovering the areas on my own and then having to figure out the tasks that I'm looking for within there. It leaves, I don't know, some to the anticipation of the game}”.  Overall, these responses suggest that discovery-based exploration fosters curiosity and deeper engagement by making challenge an intrinsically motivating part of the VR experience.

\subsubsection{\textbf{Theme 2: Importance of Navigation and Locomotion Design}}
Regardless of exploration type, navigation and locomotion techniques also shaped participants’ spatial understanding of the VR environment. Several participants reported that waypoints were particularly helpful for navigation and exploration. When asked which aspects of the system supported these tasks, P3\textsubscript{B} referred to feedback from waypoints: \textit{``When it was telling you when you're right on point''}. Similarly, P5\textsubscript{LV} with residual vision explicitly highlighted the waypoint feature: “\textit{...even like when that little yellow thing was there... This is kind of interesting. So when I heard that, it was like, okay, walk here, walk, you know what I mean? Go and then you can hear it as soon as you get there, it tells you. So that was very helpful.}” 
On the other hand, for a few participants, waypoint-based navigation, while efficient, reduced opportunities for building mental maps. As P9\textsubscript{B} explained, “\textit{I like that system if all I'm trying to do is get from point A to point B. But as far as like understanding the space, if I'm using a waypoint, I'm just following the waypoints... it makes it harder to develop a mental map of the environment.}” P9\textsubscript{B} also expressed frustration with waypoint density: “\textit{I think some of the waypoints had too many waypoints in them. there was like nine waypoints when... it was like taking two steps from one to the next.}” They suggested incorporating alternative navigation strategies, such as pinging and teleporting to a VR location. 

Locomotion design further influenced spatial perception. \softwarename{}’s joystick-based approach, which maps movement to head orientation, sometimes created a mismatch between expected and perceived motion. As P8\textsubscript{LV} described, “\textit{you'd think you'd be walking forward, but you're actually walking sideways or something like that.}” Similarly, P3\textsubscript{B} noted difficulty maintaining directional intent: “\textit{Trying to visualize yourself, trying to go up [forward] and do you go straight [head direction] or do you go [forward].}” Certain control techniques also disrupted users’ sense of orientation. For example, P9\textsubscript{LV} was disoriented by joystick's snap turning, which rotates the camera view in $45^\circ$ increments: “\textit{Like I tried using the, you know, 45 degrees snap look and I just completely lost sense of direction if I use that.}”

Together, these findings highlight the importance of designing navigation and locomotion techniques that better align with BLV users’ embodied experiences to support both efficient movement and spatial understanding.

\subsubsection{\textbf{Theme 3: Multimodal Feedback as a System for Awareness and Interaction}}
Participants consistently described spatial awareness as emerging from the coordination of multiple feedback modalities rather than any single source. Instead of relying on isolated cues, they interpreted spatial information through the interplay of audio, haptic, and verbal feedback, each contributing distinct yet complementary information. As P11\textsubscript{LV} articulated, “\textit{I'm someone that I like to use all the different types of feedback... so when you have all those different types of feedback available to you, I feel like I can get a better sense of my surroundings.}”

Directional feedback was especially critical for orientation, particularly when aligning with targets. While traditional audio beacons provided continuous spatial guidance, participants emphasized that the discrete tonal changes in responsive audio beacons enabled more precise alignment. As P9\textsubscript{B} explained, “\textit{while you're getting that left to right stereo, it can... it's harder to be like, okay, I'm at zero degrees... where then you have that tone switches, you know, okay, I'm facing exactly.}” Similarly, P6\textsubscript{LV} noted, “\textit{ And it lets me know when I was going towards the right direction... if I had to go a little bit to the right or a little bit to the left, I could kind of hear it.}”

Haptic feedback conveyed proximity and alignment. P9\textsubscript{B} described this synergy: “\textit{I mean, the audio wasn't really letting me know how far away from the waypoint I was, but then once I get close enough, then the haptic would start to kick in and have some, I like having the distance information along with the direction...}” P4\textsubscript{B} also highlighted the combined effect: “\textit{I like the, not just the noise, but the intensity of the haptic feedback when I was headed in the right direction. And like when you're right on it, it's just like bzzzzzz, it's nice.}” Similarly, P6\textsubscript{LV} highlighted liking the pointing precision using haptics: \textit{``being able to like grab stuff and be right on point with it.''} Together, these modalities reduced ambiguity: audio provided directional cues, while haptics confirmed precision and proximity for user actions.

Haptics also played a key role in conveying physical interactions such as collisions. Participants described vibration as grounding the experience in tangible events. As P5 noted, “\textit{Because it was telling me there was something in front of me.}” P8\textsubscript{LV} added that haptics reduced frustration when hitting a wall: “\textit{That's what took kind of the frustration away is like if I just heard it, I'd be like, okay, what am I hitting? Whereas with the vibration, it's like, okay, you're hitting something...}” This physical feedback not only improved spatial awareness but also reduced uncertainty during interaction.

Verbal descriptions provided important contextual scaffolding. Participants noted that text-to-speech (TTS) helped them understand not just spatial layout, but the meaning of the environment. As P8\textsubscript{LV} explained, “\textit{It gave me a better description of my surroundings... when the text-to-speech said, you're in a spaceship, it made me feel like I was in the game...}” P5\textsubscript{LV} and P11\textsubscript{LV} similarly emphasized the value of audio descriptions for understanding context and orientation. By supplying semantic and narrative information, TTS complemented sensory cues, enhancing usability and immersion.

Participants suggested improvements to make multimodal feedback more informative and context-sensitive. In particular, they wanted feedback that not only confirms actions but also helps differentiate objects through distinct audio cues in complex environments. As P11\textsubscript{LV} explained, “\textit{[I] really like when the objects make a noise themselves, like a chime or something... especially if you have multiple items close to each other, you can kind of hear the different type of items and you can change the auditory depending on what type of item it is.}” Similarly, P4\textsubscript{B} suggested providing distinct haptic feedback for different objects. These suggestions highlight the value of adaptive feedback that supports both spatial guidance and rich interaction.

\section{Discussion}
Below, we reflect on our results, discuss implications for future design and research, and outline limitations of our work.

\subsection{Reflection on Results}
Our work presents \softwarename{} and investigates two research questions.

\paragraph{\textbf{(RQ1) How do BLV users experience \softwarename{}’s discovery and multimodal navigation mechanisms?}}
Participants showed a clear preference for the \emph{discovery} exploration type (n=10/12) and \emph{responsive audio beacon with haptics} (n=10/12). Qualitative findings highlight the experiential factors underlying these preferences. Participants described \emph{discovery} as intrinsically motivating, emphasizing autonomy and the satisfaction of figuring things out independently. Many reported that \emph{discovery} increased engagement and that the added challenge enhanced the overall experience. Similarly, preferences for multimodal feedback stemmed from the complementary roles of responsive audio and haptics, where responsive audio provided directional guidance, while haptics offered confirmation of alignment, proximity, and interaction. Participants frequently described this combination as giving them a more complete understanding of their surroundings.

However, these qualitative experiences did not translate into clear differences in quantitative ratings of spatial presence or agency. 
Although \emph{discovery} and responsive audio with haptics showed slightly higher average ratings than the alternatives, these differences were not statistically significant. Several factors may explain this. First, the sample size was relatively small, and participants varied in vision condition and behavioral tendencies (e.g., task-focused vs. exploratory), as reflected in differing movement patterns. The partial eta-squared values in Table~\ref{tab:ANOVA} suggest small to medium effect sizes that may need a larger sample to detect reliably. Second, \emph{discovery} may primarily influence other aspects of user experience. 
Participants described \emph{discovery} as more game-like and curiosity-driven, which increased engagement but did not necessarily improve spatial awareness compared to a \emph{prebuilt menu}. This aligns with findings from Nair et al.~\cite{nair2024surveyor} that reported higher preference and fun for their discovery approach in a video game setting and reinforces previous research emphasizing open exploration as a key factor in gamification and enjoyment~\cite{seaborn2015gamification}. 

Similarly, while responsive audio and haptics enhanced user confidence and reduced uncertainty, the continuous vibration may not have been sufficient to impact perceived presence or agency. Supporting stronger presence may require richer, physically grounded audio-haptic feedback that more closely resembles real-world interactions. 

 Importantly, these findings should not be interpreted as evidence that discovery-based exploration or multimodal feedback are objectively superior to existing approaches. While participants consistently preferred these interactions, preference alone does not demonstrate improved performance and may be influenced by experimental biases~\cite{ma2026quantifying,dell2012yours}. Instead, our results suggest that discovery and audio-haptic feedback primarily influence experiential aspects of VR, such as perceived autonomy, engagement, confidence, and willingness to explore.

\paragraph{\textbf{(RQ2) How does free-form discovery affect task performance and workload compared to a pre-discovered environment? }}
Our results show that the exploration type did not significantly affect task completion time or perceived workload. 
However, \softwarename{}’s \emph{discovery} condition resulted in significantly longer path lengths and more \emph{wandering} behavior. 
These results suggest that participants were able to engage in more extensive exploration without compromising task efficiency. In contrast, prior work by Nair et al.~\cite{nair2024surveyor} reported significantly longer task completion times for their beam-based discovery approach (comparable sample size; n=9 participants). The efficiency gain observed in our discovery setup may stem from the effectiveness of head-based environmental scanning in \softwarename{}. 
Overall, our findings suggest that free-form discovery in VR can be supported without substantially degrading task efficiency or increasing user workload.

\subsection{Implications for Research and Design}

Our work provides three implications for designing accessible VR exploration and interactions for BLV users.

\paragraph{\textbf{Improving BLV user's mental map of VR}} Our results show that participants valued combining free-form discovery with the guided waypoint navigation. Many users enjoyed active navigation and found the autonomy and satisfaction of ``figuring things out'' as important aspects of the experience, but some reported that waypoint-based guidance and joystick locomotion did not adequately support spatial awareness or mental map formation. 
This finding is in contrast to prior work showing
that joystick-based locomotion provides high spatial knowledge for sighted users~\cite{di2021locomotion,kim2024locomotion}. 
While recent studies have examined BLV users’ preferences and performance with different locomotion techniques~\cite{ribeiro2024investigating,kreimeier2020blindwalkvr}, there remains a limited understanding of how these techniques affect spatial learning and mental map construction for BLV individuals.  

Our results also highlight the need for navigation guidance that balance efficiency with opportunities for spatial understanding for BLV users. 
Prior work has shown that environmental landmarks, intersections, and signage play a key role in mental map formation during real-world navigation~\cite{chen2025visimark,kuribayashi2023pathfinder}. In VR, these elements could be supported through lightweight landmark cues and subtle highlighting of region intersections and doorways as decision points to reinforce spatial understanding without disrupting exploration. 

The choice of navigation guidance also involves tradeoffs between efficiency and active engagement. While our waypoint-based approach prioritizes user participation and sense of control, it requires greater effort. Alternatively, systems could automatically guide users along navigation paths while conveying environmental information through audio and haptic cues, reducing navigation effort at the cost of user control and active exploration. 
Future work should investigate these tradeoffs and explore embodied, adaptive navigation techniques that better align with BLV users’ real-world strategies while complementing discovery-based exploration and supporting mental map building in VR.

\paragraph{\textbf{Evoking a sense of discovery in diverse VR environments}}
In this work, we defined discovery as the process of progressively acquiring spatial knowledge through self-directed exploration. \softwarename{} realizes this process by using head orientation as a proxy for users' spatial attention and listening direction, which are important modalities for BLV users when exploring VR environments. However, a sense of discovery involves more than making object information available; it also depends on users' attention and intentional exploration. For example, rapidly turning one's head in a crowded environment without intending to identify objects may expose many objects without evoking a genuine sense of discovery. Future systems could further support discovery by adapting to users' intentions and exploration behaviors. Rather than presenting information at a fixed level of detail, they could dynamically adjust feedback granularity based on user movement and scene complexity~\cite{kneitmix2025from}. For example, systems could provide concise summaries during rapid head movements while offering richer object-level information during slower, deliberate exploration. Such adaptive mechanisms could better support spatial awareness, agency, and a sense of control in open-ended VR environments.

\paragraph{\textbf{Rich audio-haptic feedback for improving user engagement and awareness}}
\softwarename{}’s audio-haptic feedback was designed primarily to support orientation and precise target localization, but it did not significantly enhance spatial presence or object interaction. Prior work suggests that spatial environmental sounds can improve BLV users’ immersion and presence~\cite{killough2025vrsight,ji2022vrbubble}, while haptic feedback can enhance manipulation performance, engagement, and presence in VR~\cite{maunsbach2023mediated}. However, much of the prior work focuses on a small set of predefined audio and haptic cues for BLV users~\cite{shrestha2025virtual}. 
Extending these benefits to open-ended VR environments with diverse VR sounds~\cite{jain2021taxonomy} and object interactions (e.g., tapping, pressing) remains an open challenge.  
Our findings suggest that future systems should incorporate richer, more physically grounded audio and haptic cues that respond dynamically to user actions. Research on audio design space for BLV users~\cite{guerreiro2023design}, mapping environmental sounds to vibrations~\cite{li2026sound2hap}, or converting visual information to haptics~\cite{jingu2025scene2hap} can provide a basis for rich multimodal cues in VR. 
At the same time, these cues must be carefully integrated with guidance signals to avoid sensory and cognitive overload. In our study, environmental objects did not emit any additional audio feedback, and therefore the waypoint hums did not compete with other scene sounds. However, as future VR systems incorporate richer environmental audio and object-based sound and haptic effects, navigation cues will need to be carefully balanced with contextual cues. One possible design approach is to keep navigation cues abstract (e.g., hums or buzzes) to distinguish them from environmental sounds and interaction feedback, while using adaptive audio mixing to prioritize task-relevant cues. Another is to allow users to adjust cue intensity or switch feedback modes based on context, for example, transitioning from directional guidance to object-specific audio and haptics during interaction. Such designs can enable audio and haptics to support both explicit navigation and implicit spatial awareness, ultimately enhancing BLV users' engagement and presence.

\paragraph{\textbf{Leveraging AI for Scalable Accessibility Integration in VR Development}} 
Advances in multimodal large language models (MLLMs)~\cite{vidscribe,avatar,vidcomposition} present opportunities to reduce the development effort required to integrate accessibility systems like \softwarename{} into VR environments. 
Similar to recent progress in AI-driven video accessibility~\cite{li2025videoa11y,li2026adcanvas,10.1145/3357236.3395433,dis2025maryam}, future VR systems could leverage automated scene understanding to dynamically generate descriptions and interactions.
Currently, \softwarename{} relies on manually defined regions, sections, and objects, along with authored object labels and textual descriptions. The extra annotation effort for developers pose 
an important bottleneck for VR accessibility~\cite {wang2025understanding}. 
We did not incorporate AI-driven automation~\cite{liu2024artificial} in this work for two primary reasons. First, our initial experiments with state-of-the-art models (e.g., GPT-4o) revealed inconsistencies and latency issues. For example, using AI-generated text-to-speech introduced delays and occasional failures in producing correct audio outputs, leading us to instead precompute and store audio assets. Second, our focus was on interaction design and user experience, and thus we developed environments with predefined metadata (e.g., sections, Unity tags) to ensure consistency and control. 
Future systems could leverage AI to automate key aspects of accessibility integration during the VR development phase. For instance, environments could be segmented into regions and sections using geometric heuristics (e.g., spatial layout or mesh structure), with developers refining results as needed. Similarly, advances in MLLMs could generate scene descriptions and assign semantic labels to objects based on visual and contextual cues, in line with recent work on egocentric scene understanding~\cite{kulkarni2025egovita} and VR accessibility~\cite{killough2025vrsight}. Such approaches could reduce developer burden, improve scalability, and promote more consistent accessibility practices, while still allowing human oversight and customization.

\subsection{Limitations}
Our work has five main limitations. 
 First, although our waypoint-based navigation and multimodal feedback effectively guides users toward static targets, the system does not account for dynamically moving targets. In such cases, users may continuously pursue a moving object without successfully reaching it, highlighting a need for approaches that predict a moving object's future state for navigation. 
 Second, the system’s reliance on TTS feedback, while informative, may interfere with other important auditory cues within the virtual environment and increase cognitive workload. Future work can explore ways to allow users to customize TTS output (e.g., frequency and level of detail) and introduce mechanisms such as audio docking to help BLV users manage and process multiple audio sources more effectively. 
 Third, \softwarename{} uses head orientation to determine movement direction during navigation, which can cause user confusion when the head direction diverges from joystick input. Future work could mitigate this issue by notifying users of such mismatches or exploring alternative body-based navigation techniques. 
 Fourth, while menu-based navigation provides a structured way for users to select targets, it may become less efficient as the number of regions and objects grows. Longer lists can make it harder to quickly locate desired items, potentially slowing interaction. Incorporating alternative input modalities, such as voice commands (e.g., “navigate to object X”), or enhanced filtering and search mechanisms could improve usability and scalability. 
 Finally, our study evaluated \softwarename{} with a limited number of BLV participants and VR scenes in a single-session setting. This sample size may not fully capture the diversity of vision conditions and user preferences. Also, while \softwarename{} is designed to support discovery and navigation across multiple heights and floors, our user study was limited to simple single-floor virtual layouts to ensure experimental consistency. 
 Additionally, users often adapt and refine their navigation and interaction strategies over time. As a result, it remains unclear how well our findings generalize to the broader BLV population or to longer-term, repeated use in VR environments.

\section{Conclusion}

We introduced \softwarename{}, a virtual reality system that supports discovery-based exploration and navigation for BLV users through embodied head and hand interactions combined with responsive audio and haptic feedback. Our findings show a strong user preference for discovery-driven interaction and multimodal feedback, highlighting their role in enhancing engagement and sense of agency. Notably, we found no significant impact of exploration type or feedback modality on task completion time or perceived workload, indicating that users can engage in more active, exploratory behaviors without sacrificing efficiency. 
As VR continues to grow, this work moves toward a future where immersive environments are inherently inclusive, enabling BLV users to independently explore, navigate, and fully participate in virtual experiences.

\begin{acks}
We thank SAAVI Services for the Blind for supporting participant recruitment. This research was supported by the National Eye Institute (NEI) of the National Institutes of Health (NIH) under award number R01EY034562. The content is solely the responsibility of the authors and does not necessarily represent the official views of the NIH.
\end{acks}

\bibliographystyle{ACM-Reference-Format}
\bibliography{references}

\clearpage

\end{document}